\documentclass[useAMS,usenatbib]{mn2e}
\usepackage{myaasmacros}
\usepackage{subfigure}
\usepackage{graphicx}

\def\rhodd{\rho_\mathrm{DDISC}}
\def\rhoh{\rho_\mathrm{HALO}}

\def\MWone{MW1}
\def\MWtwo{H204}
\def\MWthree{H258}
\def\MWthreedark{H258dark}

\def\MWall{MW1, H204, H258}

\def\ltsima{$\; \buildrel < \over \sim \;$}
\def\simlt{\lower.5ex\hbox{\ltsima}}   
\def\gtsima{$\; \buildrel > \over \sim \;$}
\def\simgt{\lower.5ex\hbox{\gtsima}}

\newlength{\gsize}

\newcommand\bcite[1]{\citeauthor{#1} \citeyear{#1}}

\title[A dark matter disc]{A dark matter disc in three cosmological simulations of Milky Way mass galaxies}
\author[Read et al.]{J. I. Read$^1$\thanks{E-mail: justin@physik.uzh.ch}, L. Mayer$^{1,2}$, A. M. Brooks$^4$, F. Governato$^3$ and G. Lake$^1$\\
	$^1$Institute for Theoretical Physics, University of Zurich, Winterthurerstrasse 190 8047\\
	$^2$Institute for Astronomy, Physics Department, ETH Zurich, 8093 Zurich\\
	$^3$Astronomy Dept., University of Washington, Box 351580, Seattle WA 98195-1580\\
	$^4$California Institute of Technology, M/C 130-33, Pasadena, CA 91125}

\begin{document}

\maketitle

\begin{abstract}
Making robust predictions for the phase space distribution of dark
matter at the solar neighbourhood is vital for dark matter direct
detection experiments. To date, almost all such predictions have been
based on simulations that model the dark matter alone. Here, we use
three cosmological hydrodynamic simulations of bright, disc dominated
galaxies to include the effects of baryonic matter self-consistently
for the first time. We find that the addition of baryonic physics
drastically alters the dark matter profile in the vicinity of the
Solar neighbourhood. A stellar/gas disc, already in place at high redshift, causes
merging satellites to be dragged preferentially towards the disc
plane where they are torn apart by tides. This results in an accreted dark matter disc that contributes
$\sim 0.25 - 1.5$ times the non-rotating halo density at the solar position. The dark
disc, unlike dark matter streams, is an equilibrium structure that must exist in disc galaxies that form in a hierarchical cosmology. Its low rotation lag with respect to the Earth significantly boosts WIMP capture in the Earth and Sun, boosts the annual modulation signal, and leads to distinct variations in the flux as a function of recoil energy that allow the WIMP mass to be determined.
\end{abstract}


\section{Introduction}\label{sec:intro}

The case for dark matter in the Universe is based on a wide range of observational data, from galaxy rotation curves and gravitational lensing, to the Cosmic Microwave Background Radiation (e.g. \bcite{2001ApJ...552L..23D}; \bcite{2007arXiv0704.3267R}; \bcite{2008arXiv0803.0586D}; \bcite{2006ApJ...648L.109C}). Of the many plausible dark matter candidates in extensions to the Standard Model, Weakly Interacting Massive Particles (WIMPs) stand out as well-motivated and detectable \citep{1996PhR...267..195J}, giving rise to many experiments designed to detect WIMPs in the lab (\bcite{2008EPJC..tmp..167B}; \bcite{2005NewAR..49..289A}; \bcite{2008arXiv0802.3530C}). Predicting the flux of dark matter particles through the Earth is key to the success of such experiments, both to motivate detector design, and for the interpretation of any future signal \citep{1996PhR...267..195J}. 

Most previous work has modelled the distribution of dark matter at the solar neighbourhood using, or extrapolating from, cosmological simulations that model only the dark matter (\bcite{2008arXiv0808.2981S};  \bcite{2008ApJ...686..262K}; \bcite{2008PhRvD..77j3509K}; \bcite{2008Natur.456...73S}; \bcite{2008MNRAS.385..236V}). These exquisitely describe the amount and properties of dark matter substructure. But they do not model the baryonic component of the Milky Way that presently dominates the mass interior to the solar radius (\bcite{1998MNRAS.294..429D}; \bcite{2002ApJ...573..597K}; \bcite{2005ApJ...631..838W}), and likely did so since redshift $z=1$, when the mean merger rate in a $\Lambda$CDM cosmology peaked (\bcite{2002AJ....124.1328D}; \bcite{2007ApJ...663L..13B}; \bcite{2001ApJ...546..223G}; \bcite{2006astro.ph.11187K}).

In recent work, we demonstrated that the Milky Way stellar/gas disc is important because it biases the accretion of satellites, causing them to be dragged towards the disc plane \citep{2008MNRAS.389.1041R}. As these satellites are torn apart by tidal forces, their accreted material settles into a thick disc of stars and dark matter (\bcite{2008MNRAS.389.1041R}; \bcite{1989AJ.....98.1554L}). The dark disc, unlike dark matter streams, is an equilibrium structure that must exist in disc galaxies that form in a hierarchical cosmology. Its low rotation lag with respect to the Earth significantly boosts WIMP capture in the Earth and Sun (\bcite{2008arXiv0811.4172B}; \bcite{bruchinprep}), boosts the annual modulation signal, and leads to distinct variations in the flux as a function of recoil energy that allow the WIMP mass to be determined \citep{2008arXiv0804.2896B}.

In \citet{2008MNRAS.389.1041R}, we used dark matter only simulations to quantify the expected merger history for Milky Way mass galaxies in $\Lambda$CDM, and a series of isolated collisionless satellite-Milky Way merger simulations to estimate the expected properties of the dark disc. We found that, for plausible merger histories, the dark disc contributes $\sim 0.25 - 1$ times the halo density at the solar neighbourhood. In this paper, we make the first attempt to include the baryonic matter self-consistently, using three $\Lambda$CDM cosmological hydrodynamic simulations of Milky Way mass galaxies. Both approaches are complementary in quantifying the expected properties of the dark disc. Our previous approach allowed us to specify precisely the properties of the Milky Way disc and its merging satellites at high redshift; our current approach is fully self-consistent. We model the radiative cooling, star formation and feedback physics that lead to the formation of realistic disc galaxies in their cosmological context. Recent improvements in such simulations have made it possible to form galaxies with a significant disc component, and reproduce some basic galaxy properties over a range of redshifts (\bcite{2008MNRAS.387..364Z}; \bcite{2008arXiv0812.0976S}; \bcite{2007MNRAS.374.1479G}; \bcite{2008MNRAS.390.1349P}). However, due to the stochastic nature of galaxy assembly in $\Lambda$CDM, and our limited sample of simulated galaxies, we do not attempt to form genuine Milky Way analogues. Instead, we choose three galaxies with mass comparable to that of the Milky Way that span a range of interesting assembly histories. We infer from these plausible dark disc properties for the Milky Way. 

This paper is organised as follows. In \S\ref{sec:simulations}, we describe the state-of-the-art cosmological hydrodynamic simulations. In \S\ref{sec:results}, we present our results. In \S\ref{sec:discussion} we discuss our numerical limitations and the implications of our results for the Milky Way. Finally, in \S\ref{sec:conclusions}, we present our conclusions. 

\section{The cosmological hydrodynamic simulations}\label{sec:simulations}

We use three cosmological hydrodynamic simulations of Milky Way mass galaxies, two of which (\MWone,\MWthree) have already been presented in \cite{2007ApJ...655L..17B}, \cite{2008arXiv0812.0007B}, \cite{2008arXiv0801.3845M}, \cite{2008ASPC..396..453G}, and \cite{2008arXiv0812.0379G}. All three were run with the GASOLINE code \citep{2004NewA....9..137W} using ``blastwave feedback" \citep{2006MNRAS.373.1074S}; the simulation labels, parameters and choice of cosmology are given in Table \ref{tab:simulations}. We also ran a fourth simulation, \MWthreedark. This had the same initial conditions as \MWthree, but was run with only dark matter particles, and at slightly lower mass resolution. (Note that since the matter density of the Universe $\Omega_m$ is fixed, the dark matter density is higher in \MWthreedark\ than in \MWthree.) The final outputs were mass and momentum centred using the `shrinking sphere' method described in \cite{2006MNRAS.tmp..153R}, and rotated into their moment of inertia eigenframe with the $z$ axis perpendicular to the disc. For \MWone, \MWtwo\ and \MWthree, the eigenframe was found using the stars, for \MWthreedark\ it was found using the dark matter. 

The three galaxies \MWone, \MWtwo\ and \MWthree\ were chosen to span a range of interesting merger histories, rather than as perfect Milky Way analogues. \MWone\ had a very quiescent merger history, with no major mergers after redshift $z=2$; \MWtwo\ had several massive mergers after redshift $z=1$; and \MWthree\ was an extreme case with a very massive $\sim$1:1 merger at $z=1$. In \S\ref{sec:discussion}, we discuss which of these is most like our own Galaxy. 

\begin{table*}
\begin{center}
\setlength{\arrayrulewidth}{0.5mm}
\begin{tabular}{llllll}
\hline
Simulation & $(\Omega_{\rm m},\Omega_{\Lambda},\sigma_8,h)$ & $(N_{dm},N_*,N_{gas})/10^6$ & min$(M_{dm},M_*,M_{gas})/10^5\,$M$_\odot$ & $\epsilon_{dm,*,gas}/$kpc & $M_{dm}^{<300\mathrm{kpc}}/10^{12}\,$M$_\odot$\\
\hline
\MWone & $(0.3,0.7,0.9,0.7)$ & $(2.8,3.1,1.5)$ & $(7.6,0.2,0.3)$ & $0.3$ & 1.1\\
\MWtwo & $(0.24,0.76,0.77,0.73)$ & $(4,3.3,1.7)$ & $(10.1,0.41,0.58)$ & $ 0.35$ & 0.8\\
\MWthree & $(0.24,0.76,0.77,0.73)$ & $(3.5,2.2,1.4)$ & $(10.1,0.35,0.58)$ & $ 0.35$ & 0.75\\
\MWthreedark & $(0.24,0.76,0.77,0.73)$ & $(3.5,-,-)$ & $(12.25,-,-)$ & $ 0.35$ & 0.9\\

\hline
\end{tabular}
\end{center}
\caption[]{Simulation labels and parameters. From left to right the columns show the simulation label, the cosmological parameters used, the number of dark, star and gas particles at redshift $z=0$, the minimum dark matter, star and gas particle masses at $z=0$, the dark matter, star and gas force softenings (these are always equal), and the dark matter mass within 300\,kpc at $z=0$. \MWthreedark\ was set up with the same initial conditions as \MWthree, but run with only dark matter particles, and at slightly lower mass resolution.}
\label{tab:simulations}
\end{table*}

The analysis was performed as in \cite{2008MNRAS.389.1041R}. The subhalos inside each `Milky Way' and at each redshift output were identified using the Amiga Halo Finder (AHF\footnote{{\tt http://www.aip.de/People/AKnebe/AMIGA/}}) algorithm \citep{2004MNRAS.351..399G}. The subhalos were then traced backwards in time, the progenitor of each being the subhalo at the previous redshift that contains the majority of its particles. To avoid ambiguities in this halo tracking, we ordered subhalos by mass so that subhalos were linked to their most massive progenitor not already claimed by a larger subhalo. A final complication can occur if two subhalos are about to merge. In this situation AHF sometimes over-estimates the mass of the smaller of the two. We dealt with this problem by searching for sudden spikes in mass at pericentre and removing these by assigning instead the mass found at the previous output time. We define a subhalo as merged (disrupted) if it has less than a tenth of its peak circular speed considered over all times; our results are not sensitive to this parameter.

\begin{center}
\begin{figure*}
	\setlength{\gsize}{0.33\textwidth}
	\setlength{\subfigcapskip}{-1.02\gsize}

	\subfigure[\MWone\hspace{0.4\gsize}]
	{
		\label{fig0a}
		\includegraphics[height=\gsize]{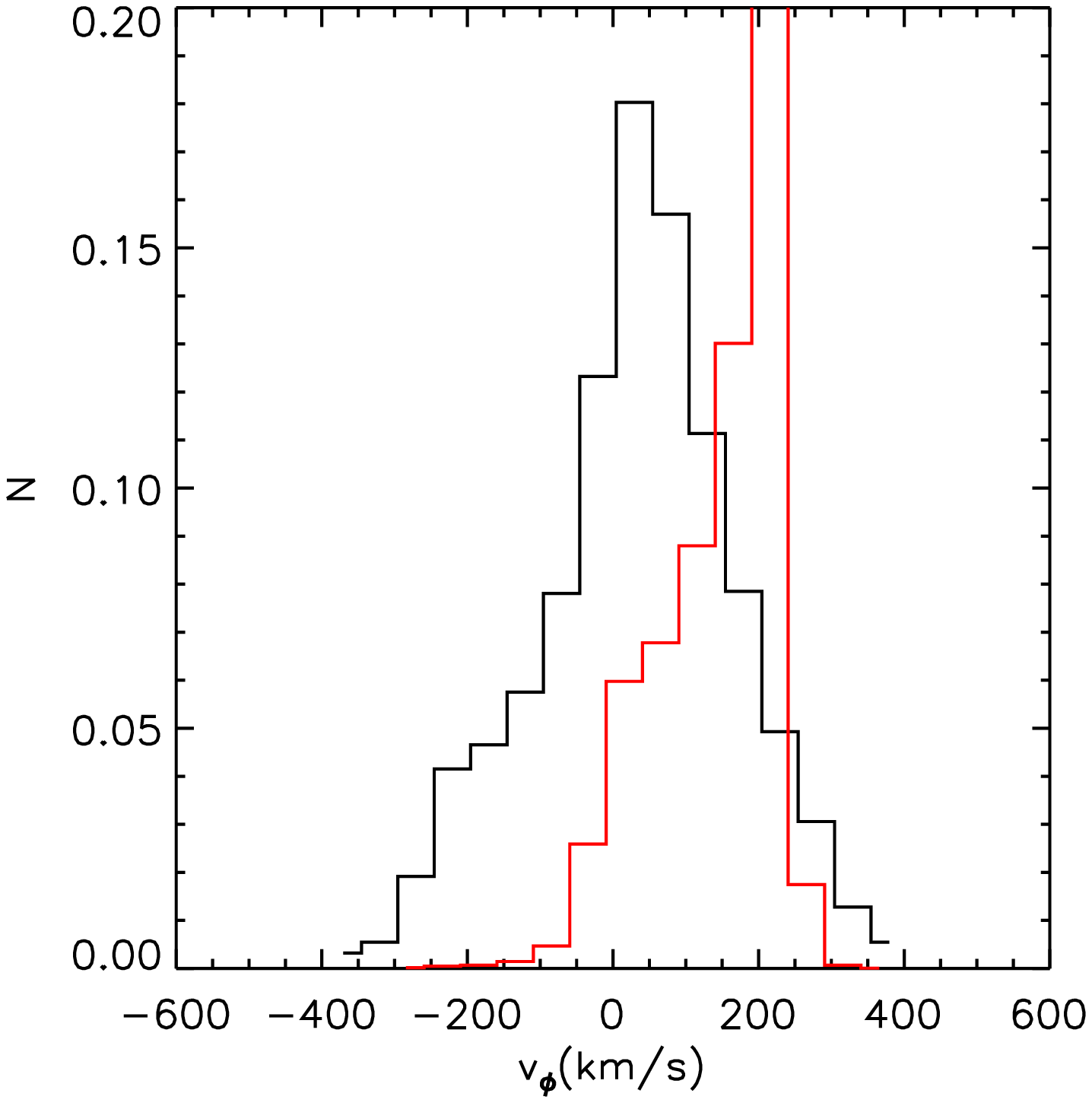}
	}
	\hspace{-0.15\gsize}
	\subfigure[\MWtwo\hspace{0.4\gsize}]
	{
		\label{fig0b}
		\includegraphics[height=\gsize]{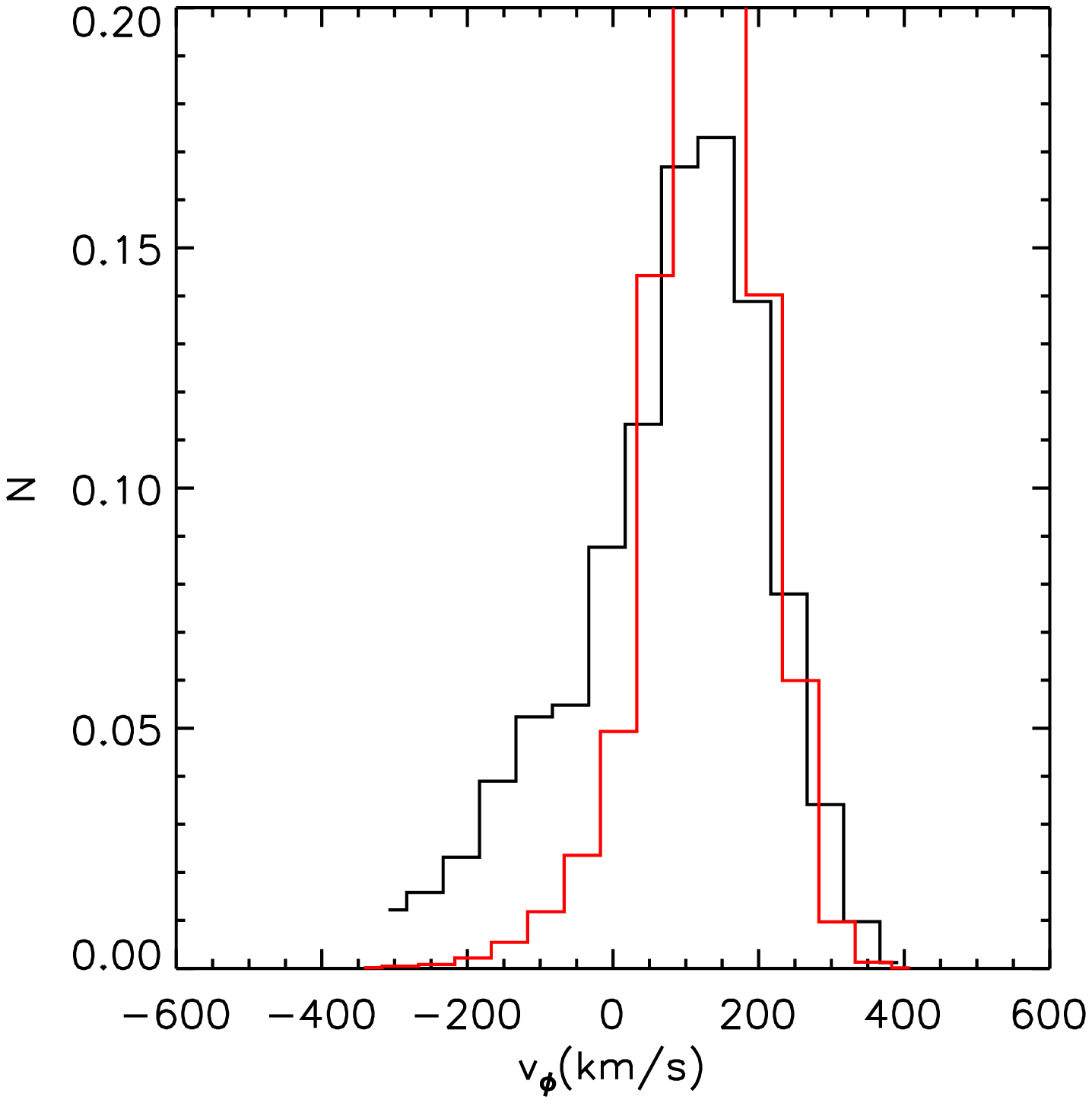}
	}
	\hspace{-0.15\gsize}
	\subfigure[\MWthree/\MWthreedark\hspace{0.15\gsize}]
	{
		\label{fig0c}
		\includegraphics[height=\gsize]{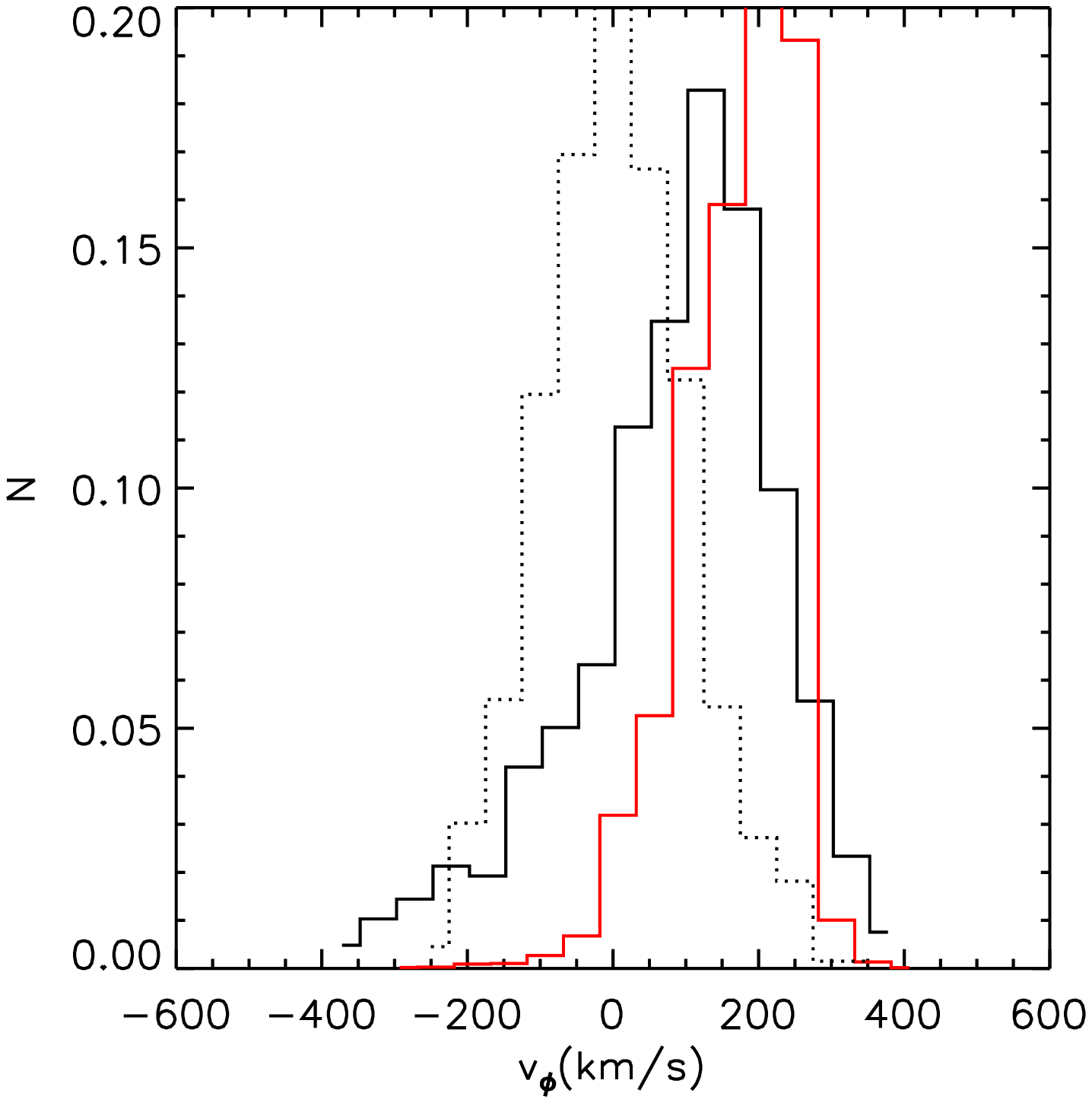}
	}
\caption{{\bf (a-c)} The distribution of rotational velocities at the solar neighbourhood ($7<R<8$\,kpc; $|z|<2.1$\,kpc) for three simulated Milky Way mass galaxies \MWone, \MWtwo\ and \MWthree. The lines show the dark matter (black) and stars (red). The dark matter distribution for \MWthreedark, simulated with {\it dark matter alone}, is overplotted on (c) (black dotted).}
\label{fig:precosmo}
\vspace{-3mm}
\end{figure*}
\end{center}

\begin{center}
\begin{figure*}
	\setlength{\gsize}{0.27\textwidth}
	\setlength{\subfigcapskip}{-1.05\gsize}

	\subfigure[\MWone\hspace{0.35\gsize}]
	{
		\label{fig1a}
		\includegraphics[height=\gsize]{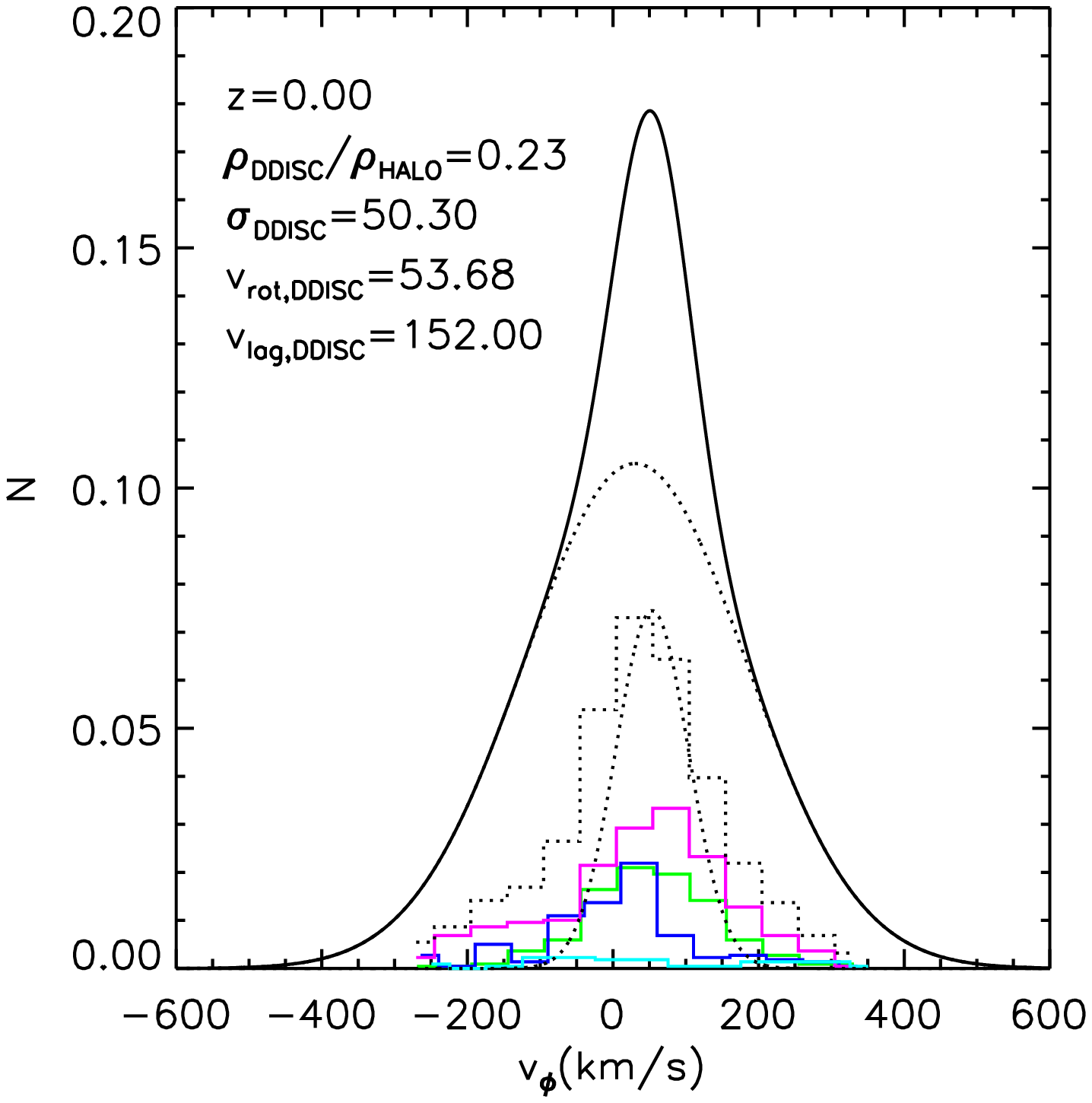}
	}
	\hspace{-0.2\gsize}
	\subfigure[\MWtwo\hspace{0.35\gsize}]
	{
		\label{fig1b}
		\includegraphics[height=\gsize]{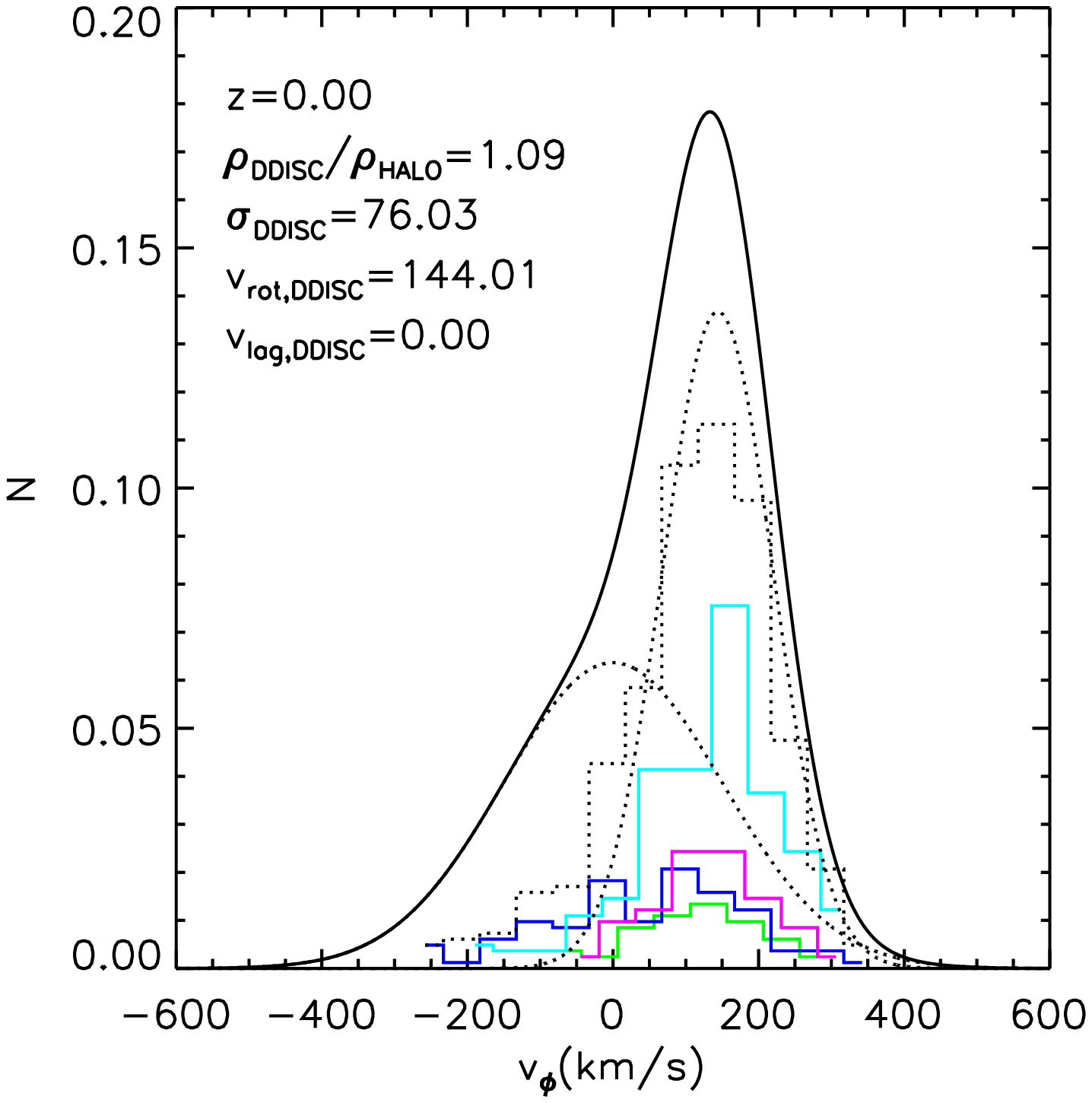}
	}
	\hspace{-0.2\gsize}
	\subfigure[\MWthree\hspace{0.35\gsize}]
	{
		\label{fig1c}
		\includegraphics[height=\gsize]{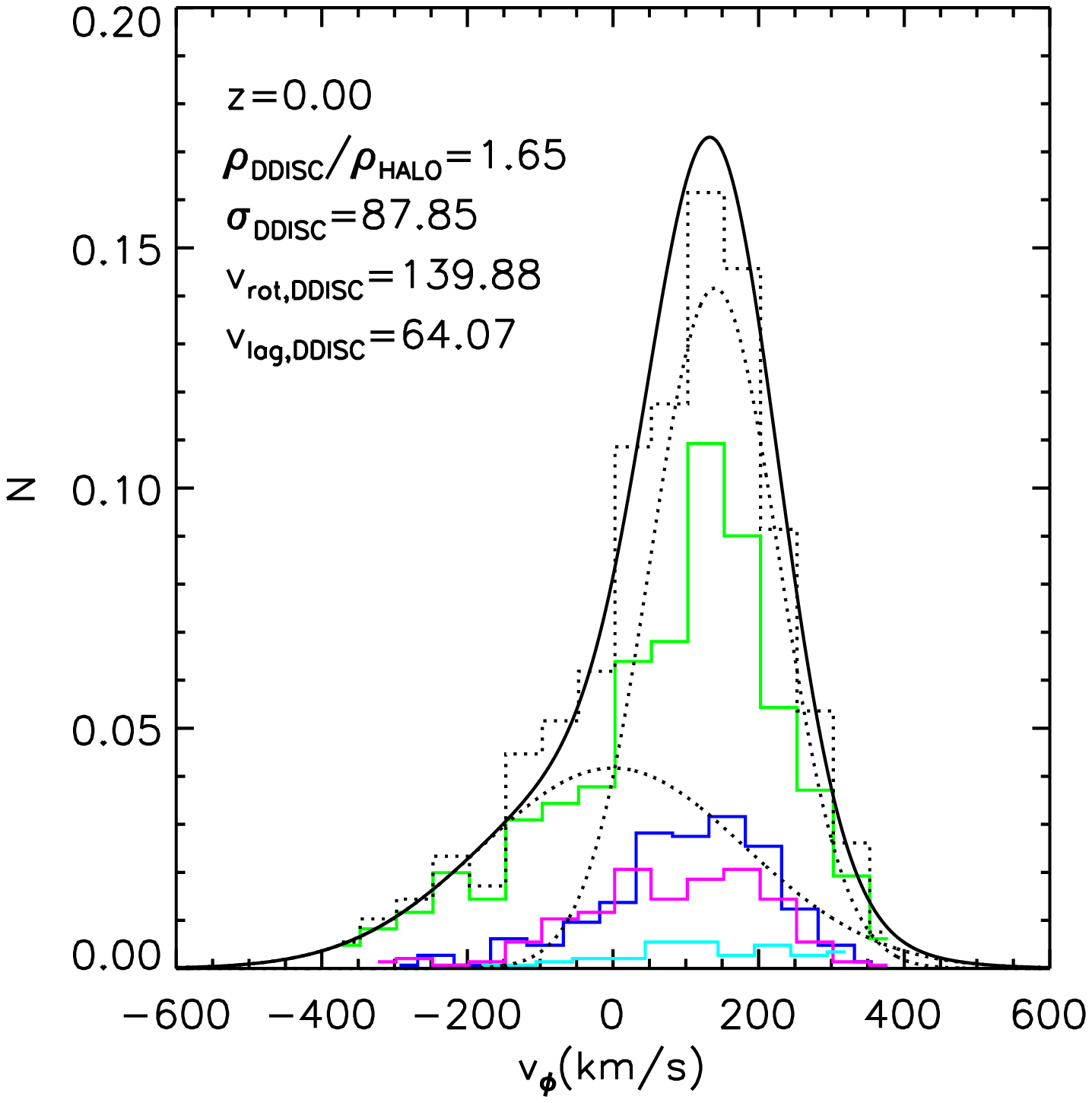}
	}
	\hspace{-0.2\gsize}
	\subfigure[\MWthreedark\hspace{0.22\gsize}]
	{
		\label{fig1d}
		\includegraphics[height=\gsize]{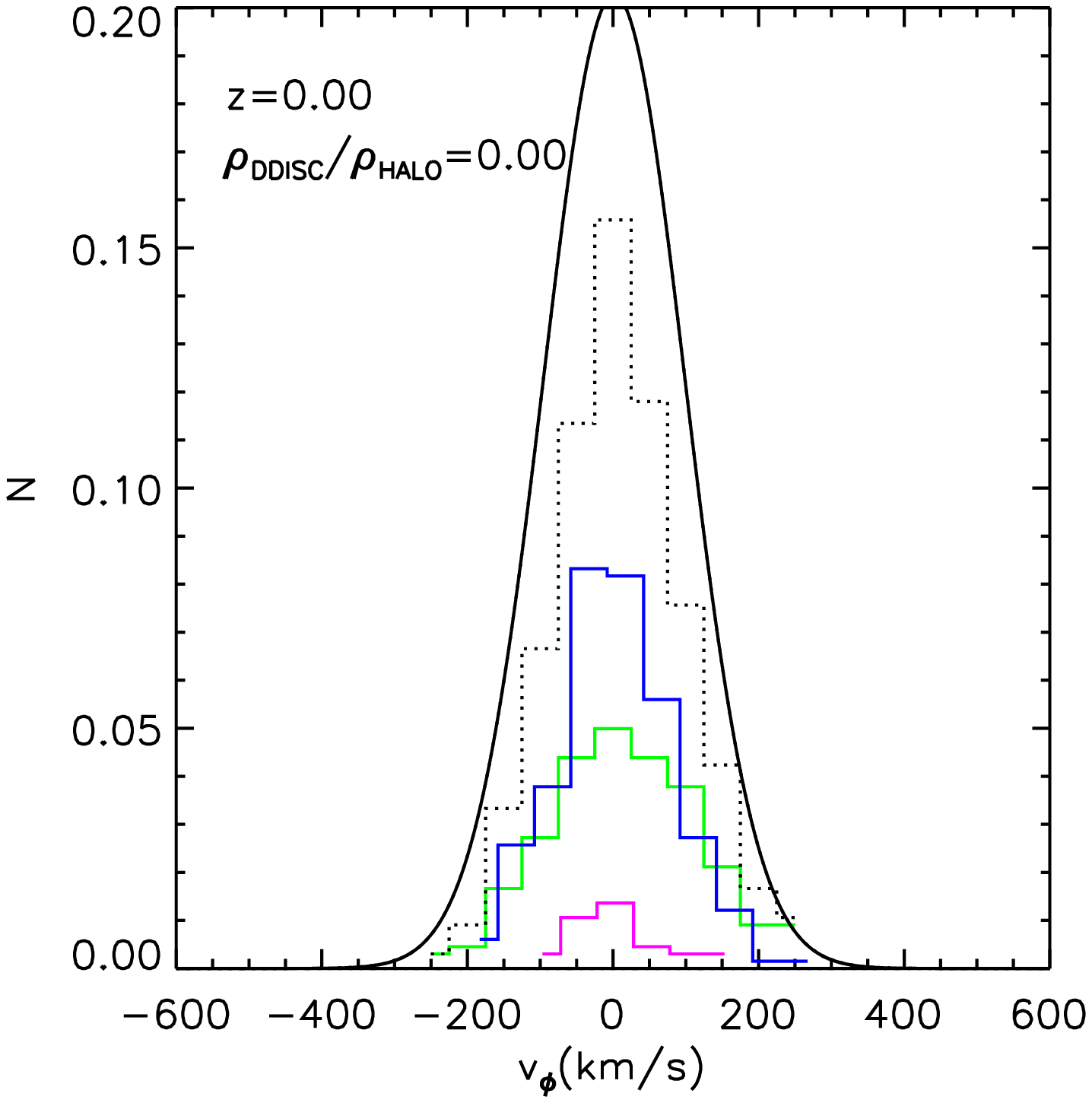}
	}\\
	
	\subfigure[\MWone\hspace{0.35\gsize}]
	{
		\label{fig1m}
		\includegraphics[height=\gsize]{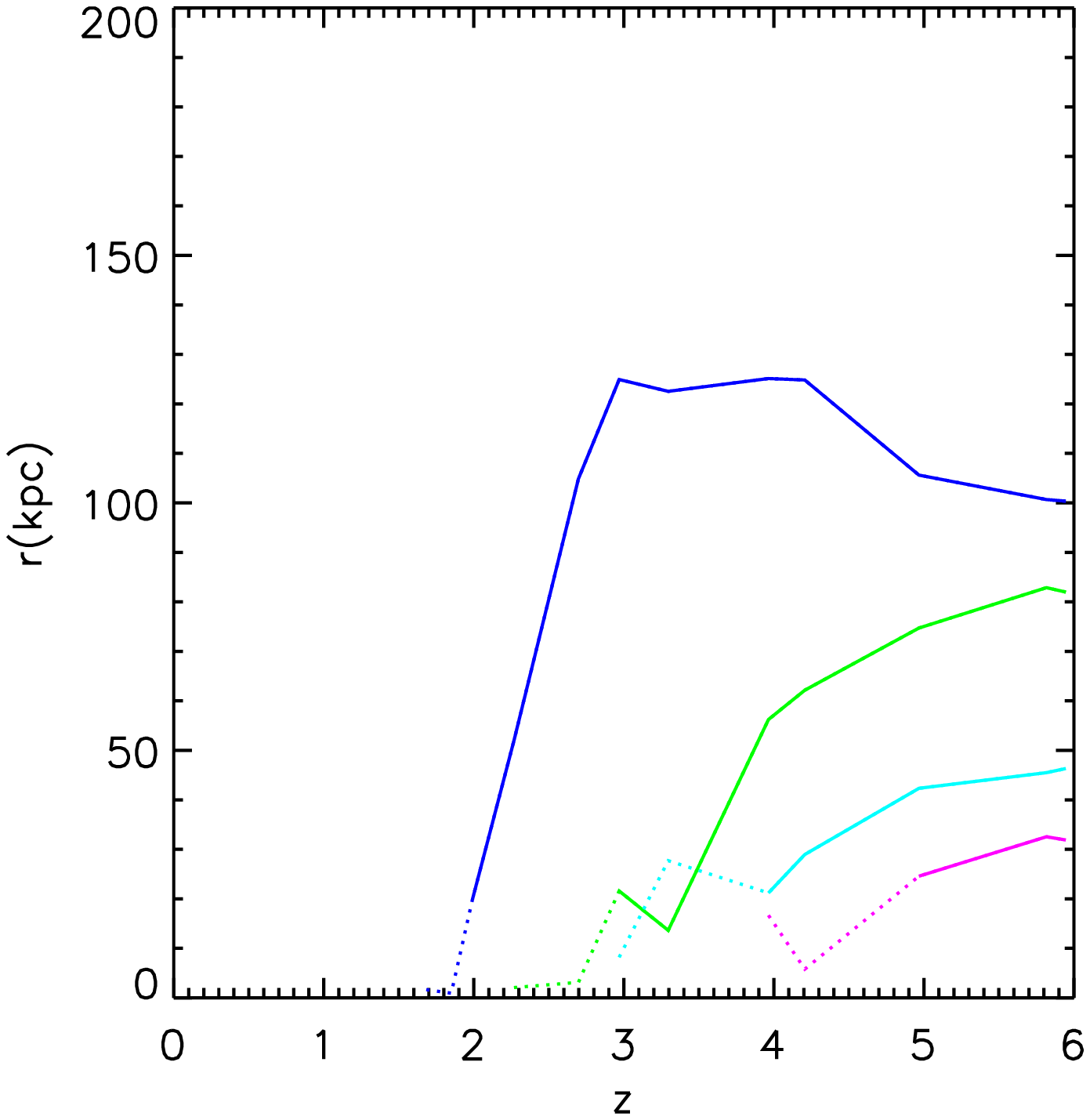}
	}
	\hspace{-0.2\gsize}
	\subfigure[\MWtwo\hspace{0.35\gsize}]
	{
		\label{fig1n}
		\includegraphics[height=\gsize]{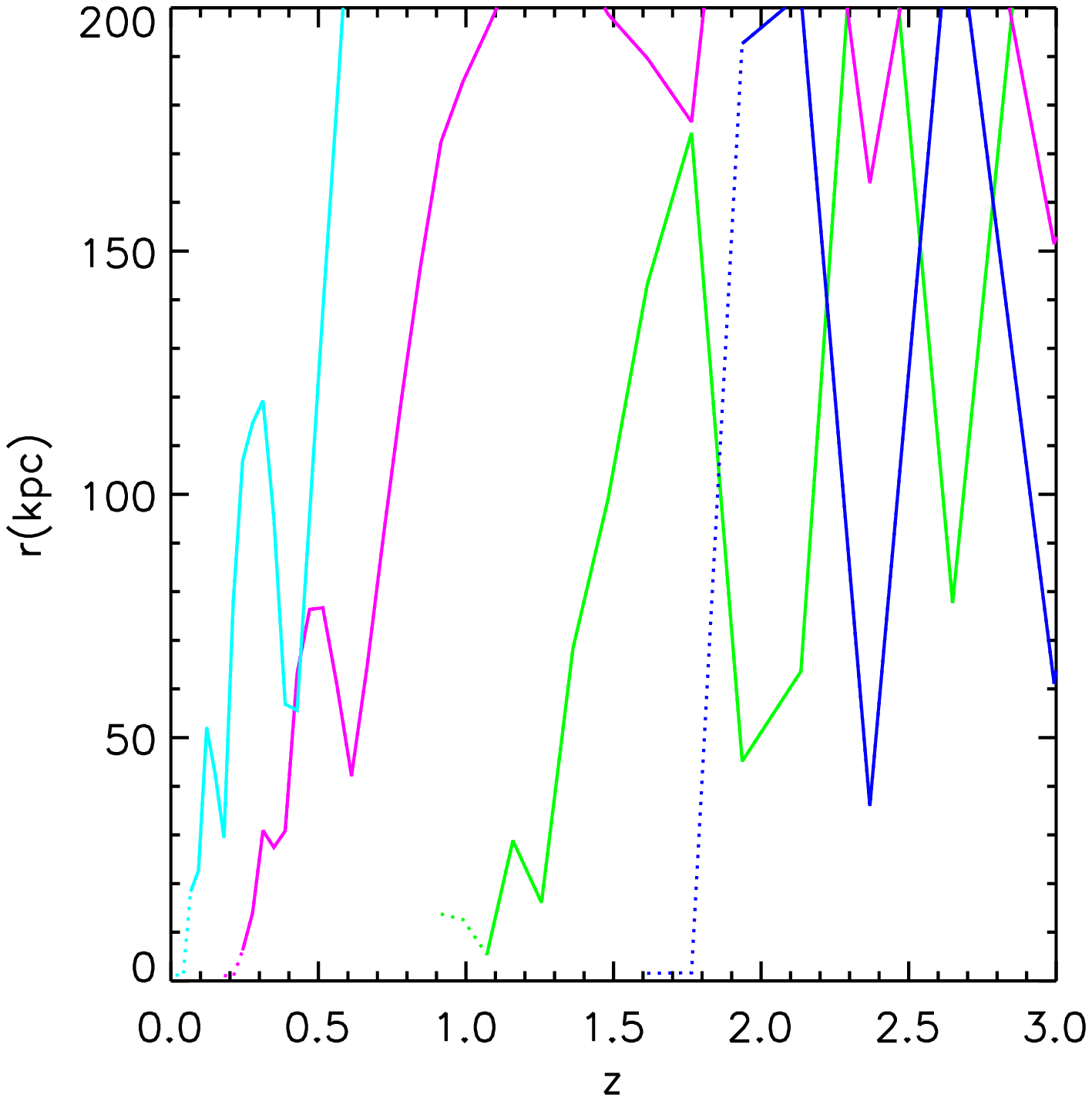}
	}
	\hspace{-0.2\gsize}
	\subfigure[\MWthree\hspace{0.35\gsize}]
	{
		\label{fig1o}
		\includegraphics[height=\gsize]{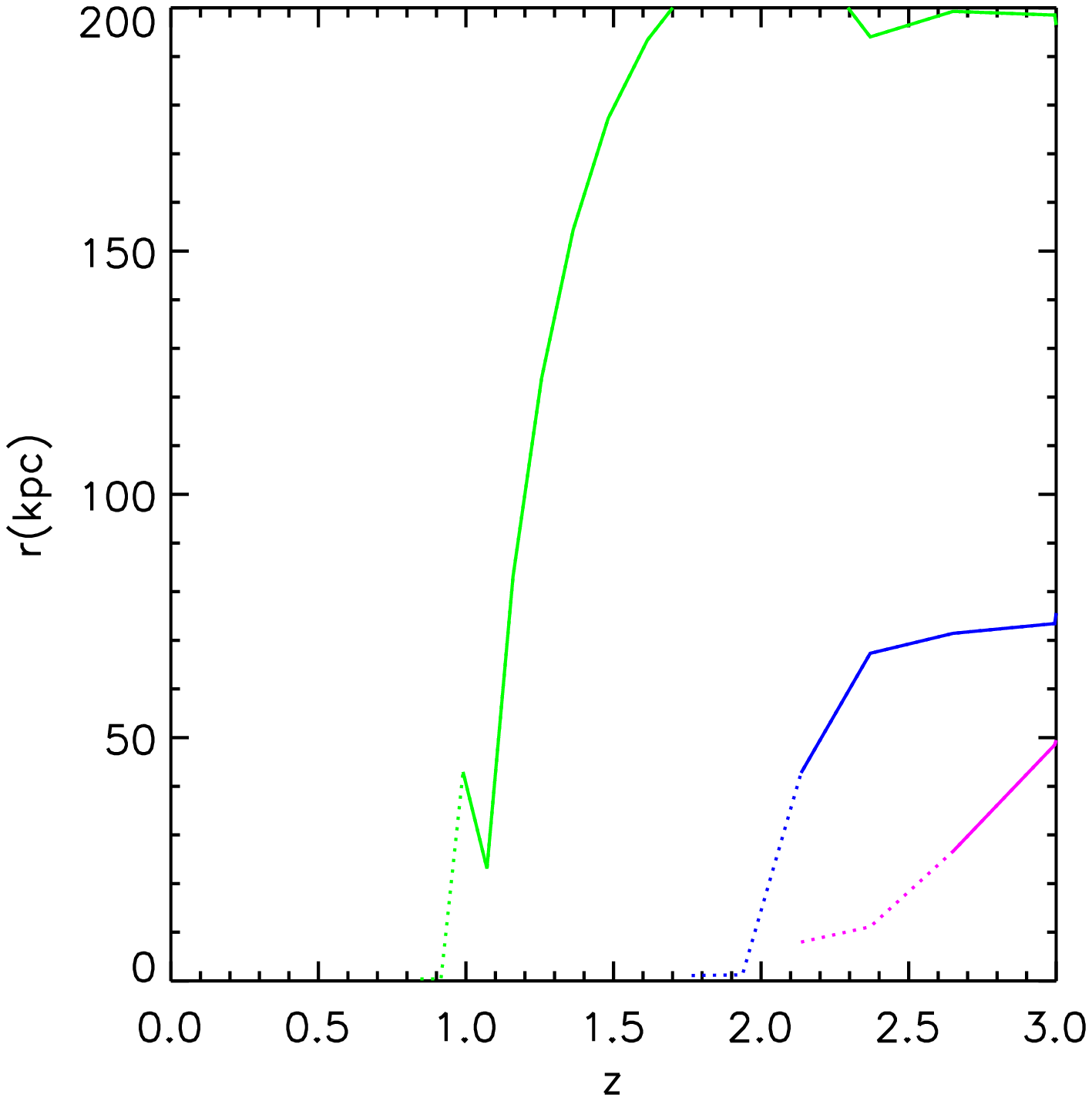}
	}
	\hspace{-0.2\gsize}
	\subfigure[\MWthreedark\hspace{0.22\gsize}]
	{
		\label{fig1p}
		\includegraphics[height=\gsize]{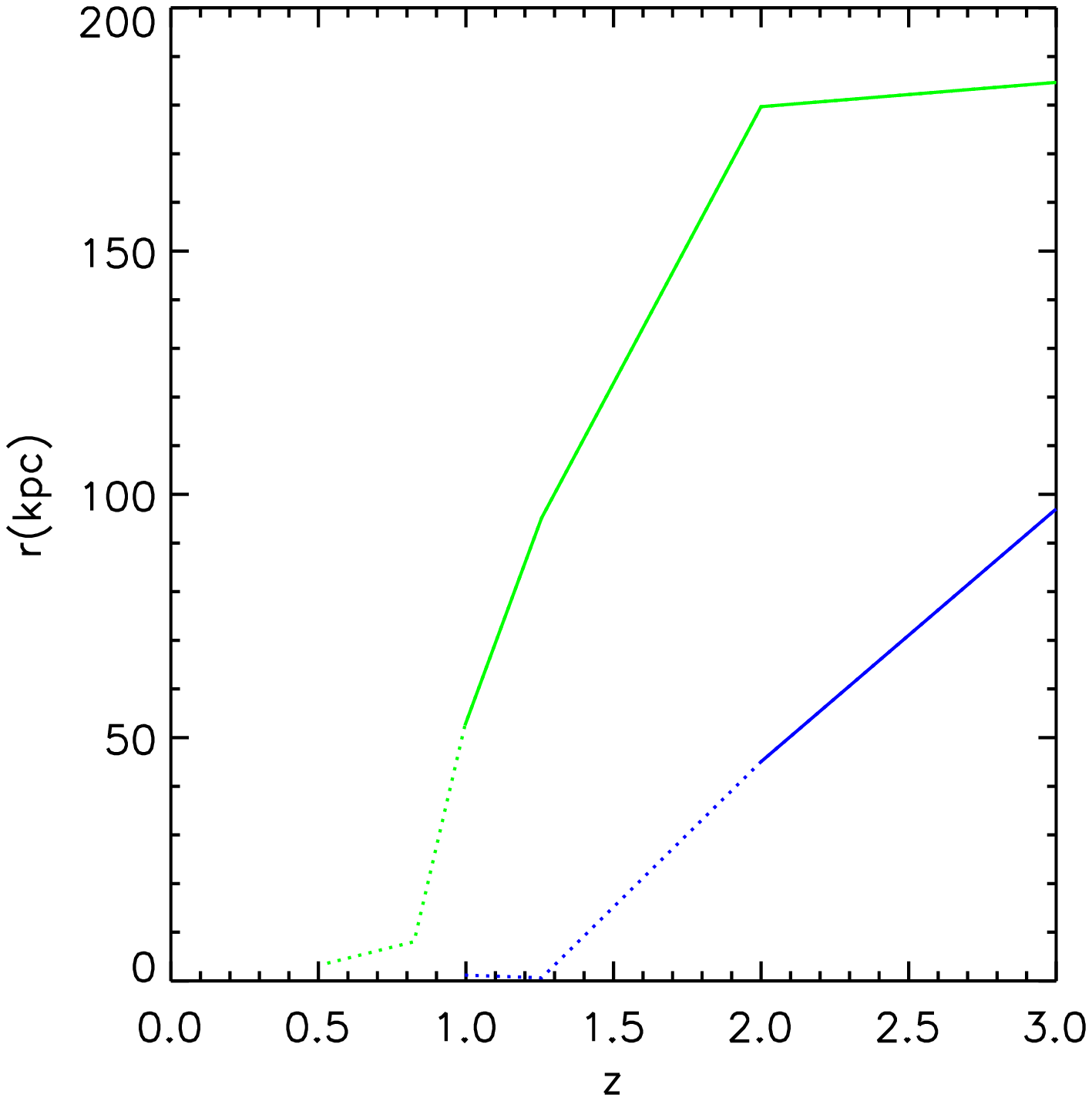}
	}\\

	\subfigure[\MWone\hspace{0.35\gsize}]
	{
		\label{fig1q}
		\includegraphics[height=\gsize]{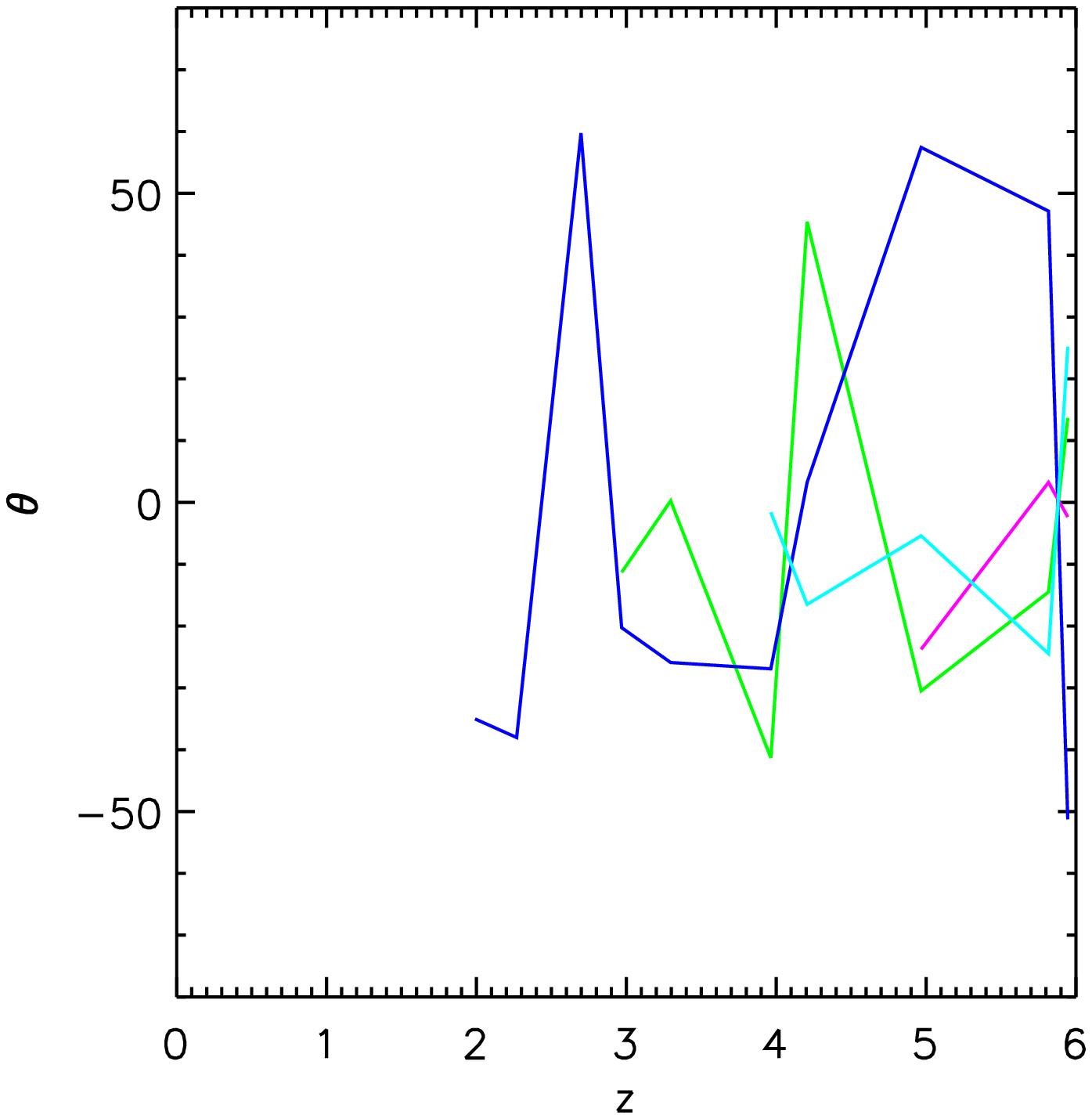}
	}
	\hspace{-0.2\gsize}
	\subfigure[\MWtwo\hspace{0.35\gsize}]
	{
		\label{fig1r}
		\includegraphics[height=\gsize]{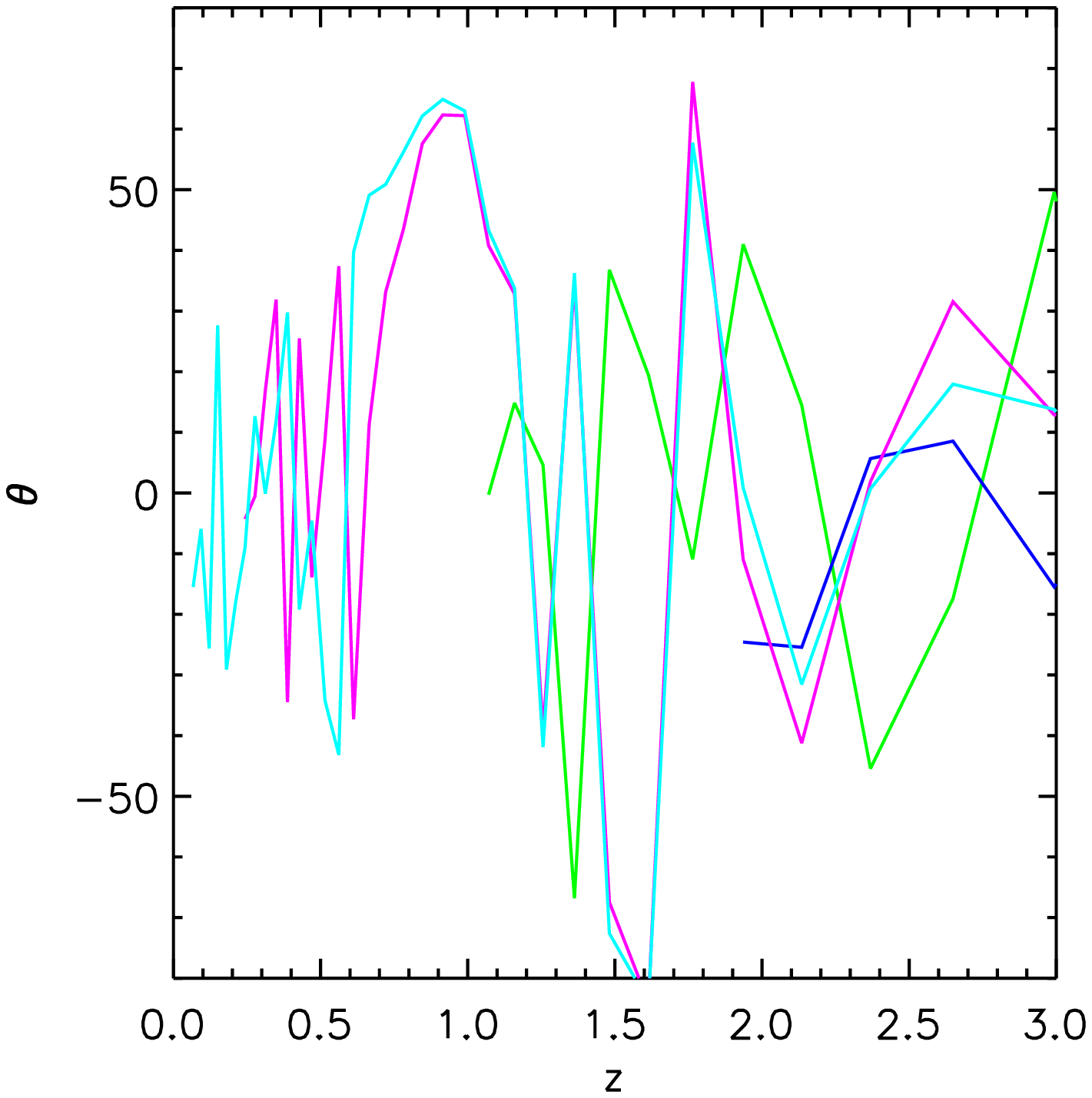}
	}
	\hspace{-0.2\gsize}
	\subfigure[\MWthree\hspace{0.35\gsize}]
	{
		\label{fig1s}
		\includegraphics[height=\gsize]{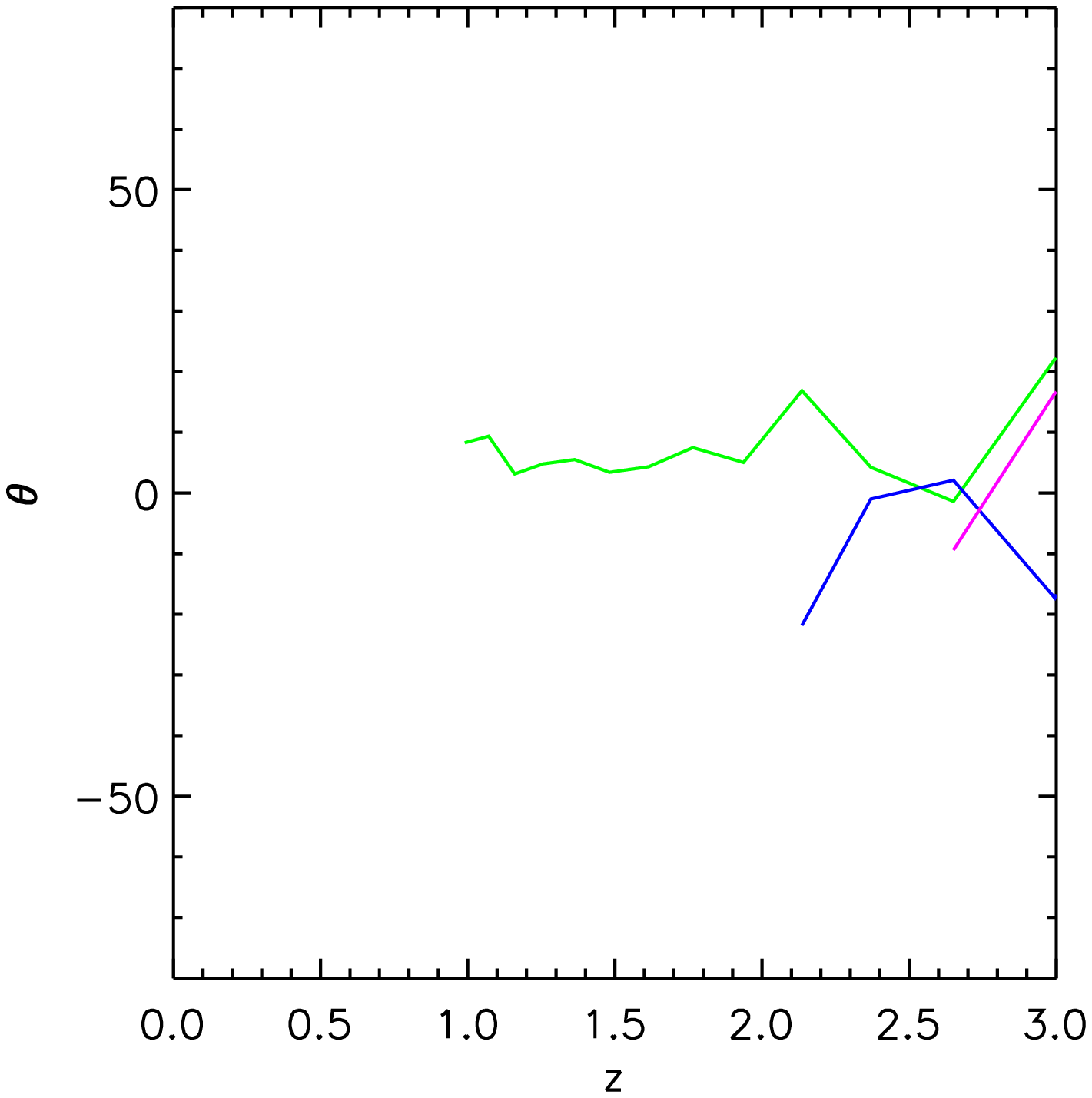}
	}
	\hspace{-0.2\gsize}
	\subfigure[\MWthreedark\hspace{0.22\gsize}]
	{
		\label{fig1t}
		\includegraphics[height=\gsize]{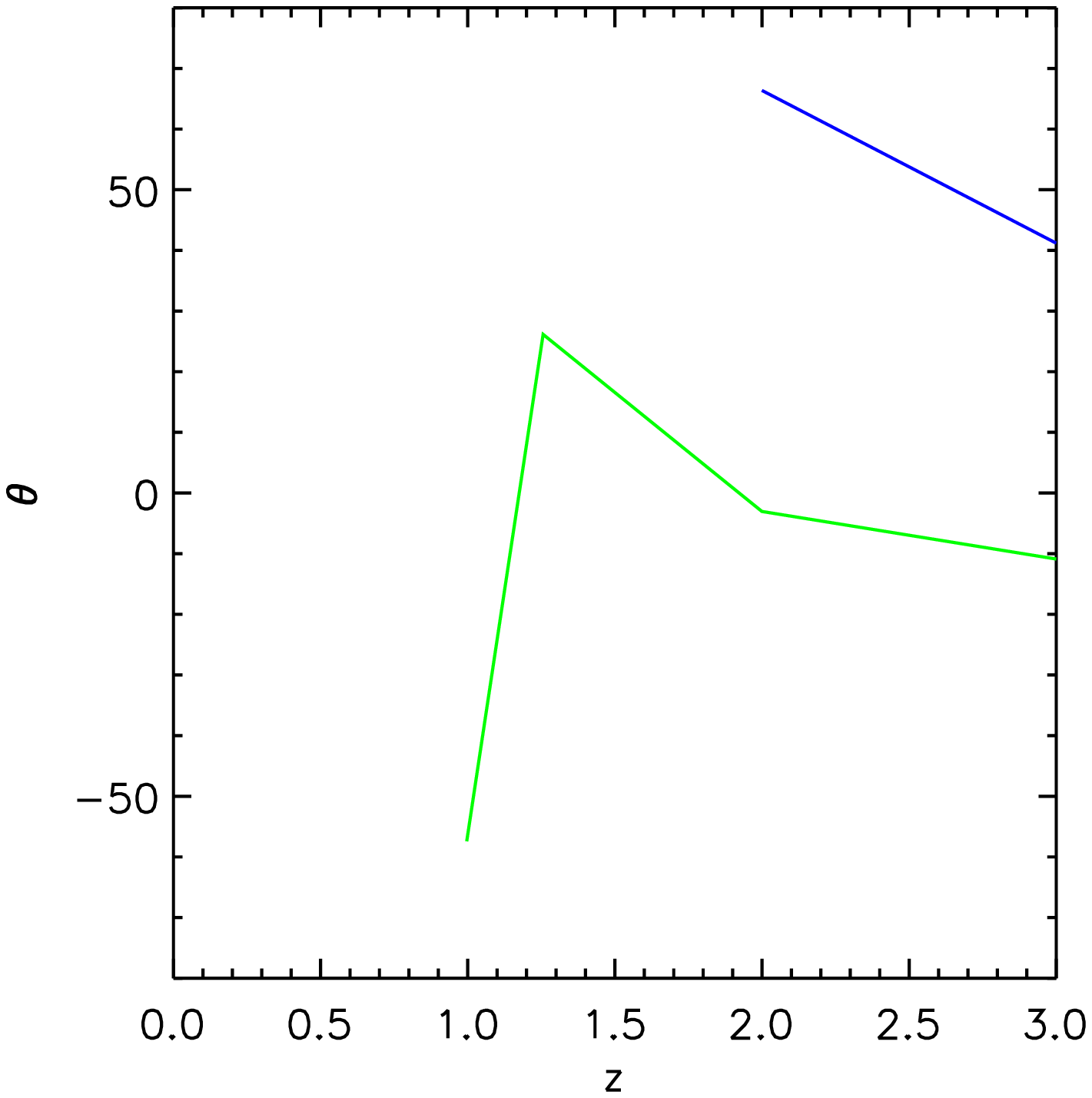}
	}\\	
\caption{{\bf (a-c)} The decomposed distribution of rotational velocities at the solar neighbourhood ($7<R<8$\,kpc; $|z|<2.1$\,kpc) for three simulated Milky Way mass galaxies \MWone, \MWtwo\ and \MWthree. The lines show a double Gaussian fit to the dark matter (black; black dotted), the dark matter accreted from the four most massive disrupted satellites (green, blue, magenta, cyan), and the sum of all dark matter accreted from these satellites (black dotted histogram). The best fit double Gaussian parameters are marked in the top left, along with the redshift, $z$. {\bf (d)} As (a-c), but for the galaxy \MWthree\ simulated with {\it dark matter alone}. {\bf (e-h)} The decay in radius $r$ as a function of redshift $z$ of the four most massive disrupting satellites in \MWall\ and \MWthreedark. Where less than four lines are shown, these satellites accreted at redshift $z>3$.  The dotted sections show the evolution of the most bound particle in the satellite after the satellite has disrupted. {\bf (i-l)} The decay in angle to the Milky Way stellar disc, $\theta$ as a function of redshift $z$ of the four most massive disrupting satellites in \MWall\ and \MWthreedark.}
\label{fig:cosmo}
\vspace{-3mm}
\end{figure*}
\end{center}

\begin{center}
\begin{figure*}
	\setlength{\gsize}{0.27\textwidth}
	\setlength{\subfigcapskip}{-1.05\gsize}

	\subfigure[\MWone\hspace{0.35\gsize}]
	{
		\label{fig1e}
		\includegraphics[height=\gsize]{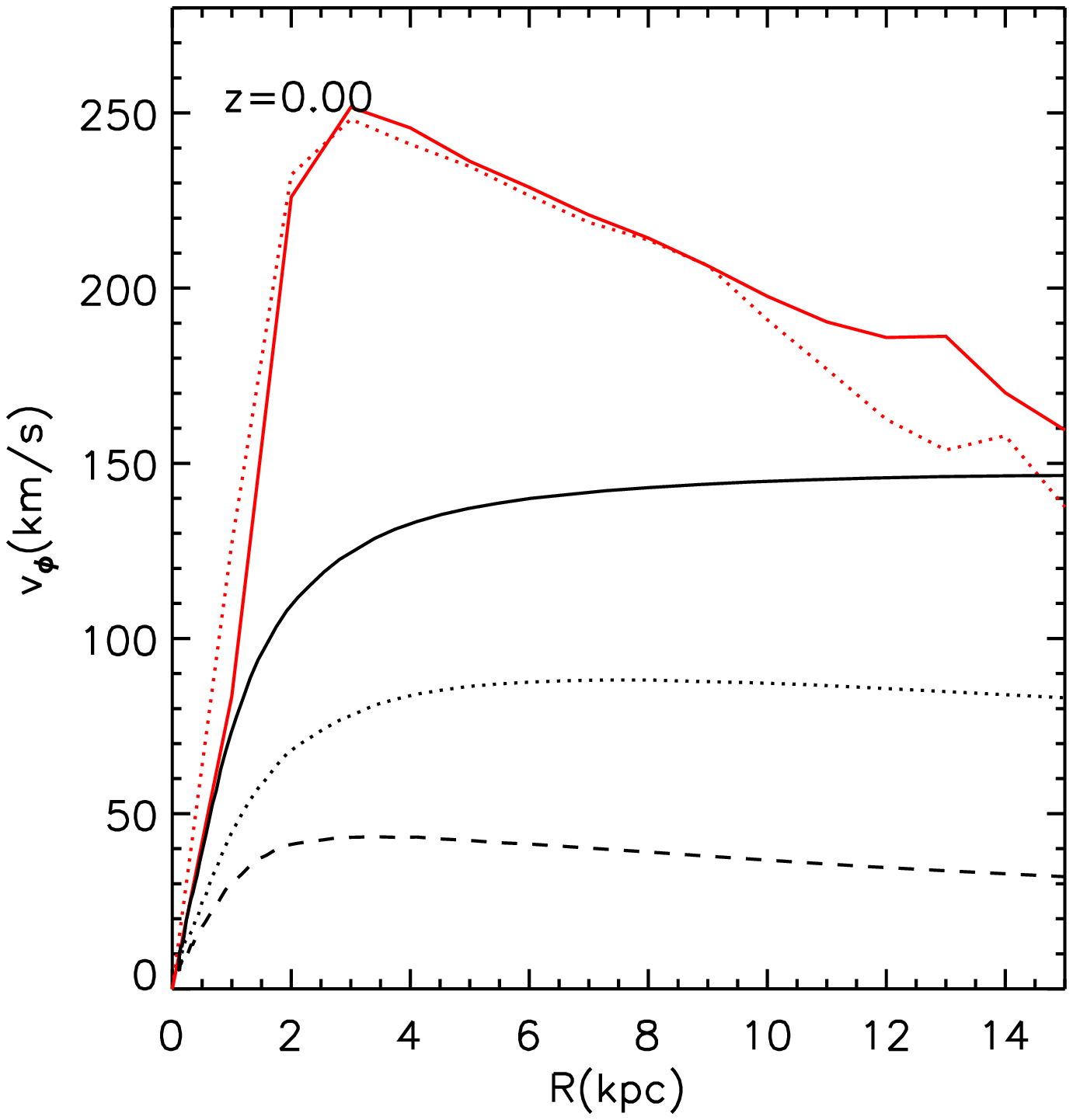}
	}
	\hspace{-0.2\gsize}
	\subfigure[\MWtwo\hspace{0.35\gsize}]
	{
		\label{fig1f}
		\includegraphics[height=\gsize]{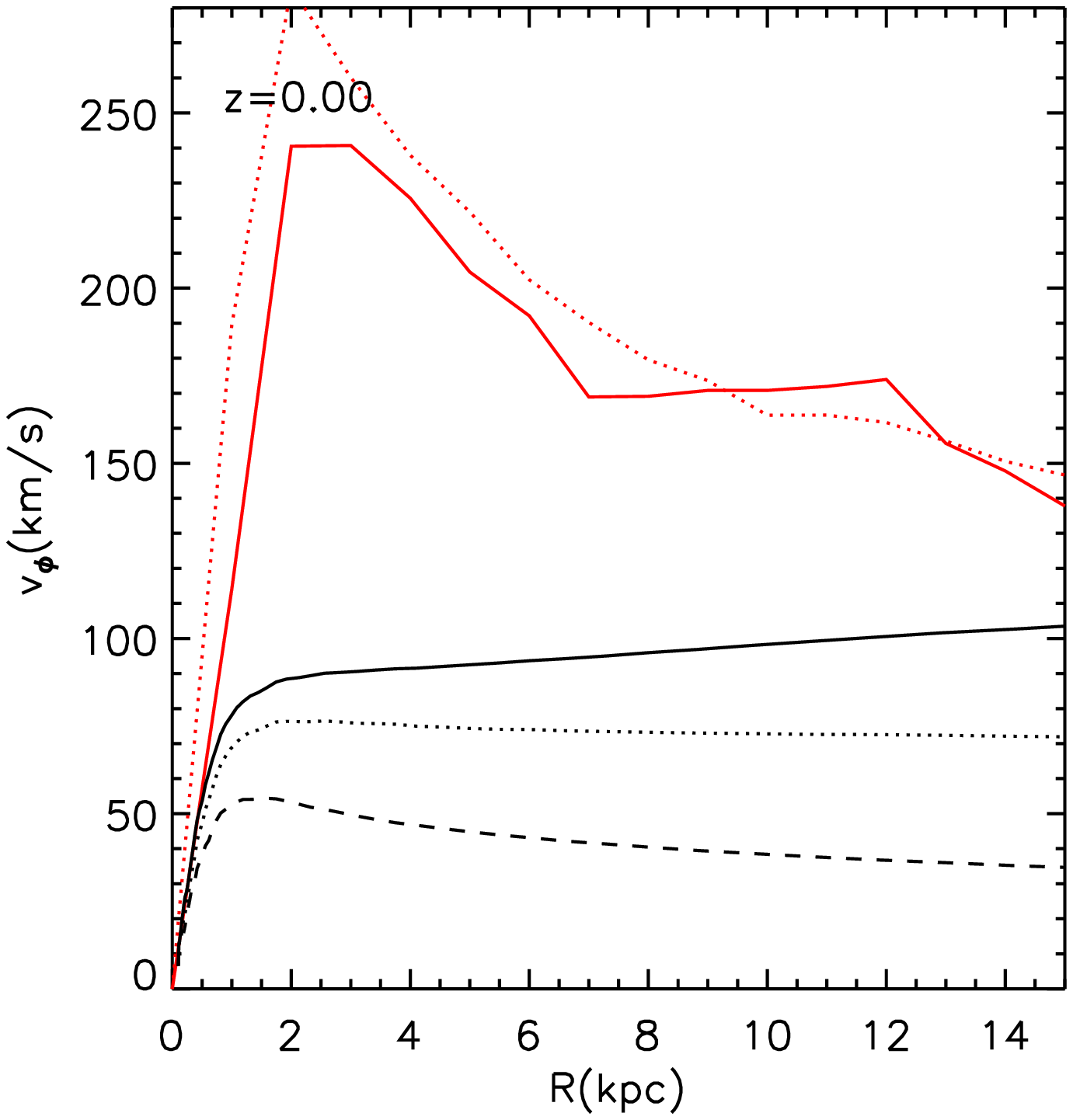}
	}
	\hspace{-0.2\gsize}
	\subfigure[\MWthree\hspace{0.35\gsize}]
	{
		\label{fig1g}
		\includegraphics[height=\gsize]{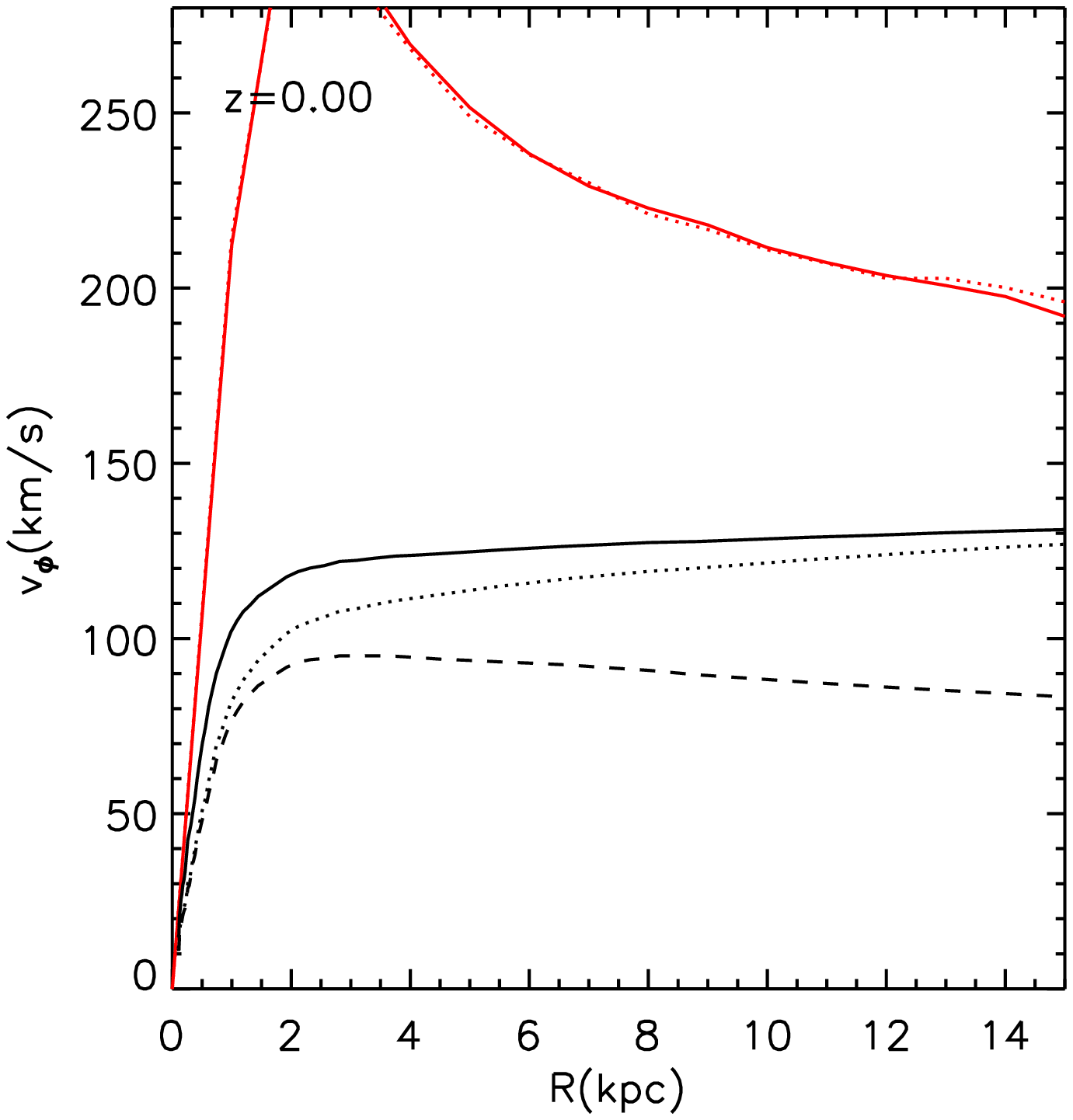}
	}
	\hspace{-0.2\gsize}
	\subfigure[\MWthreedark\hspace{0.22\gsize}]
	{
		\label{fig1h}
		\includegraphics[height=\gsize]{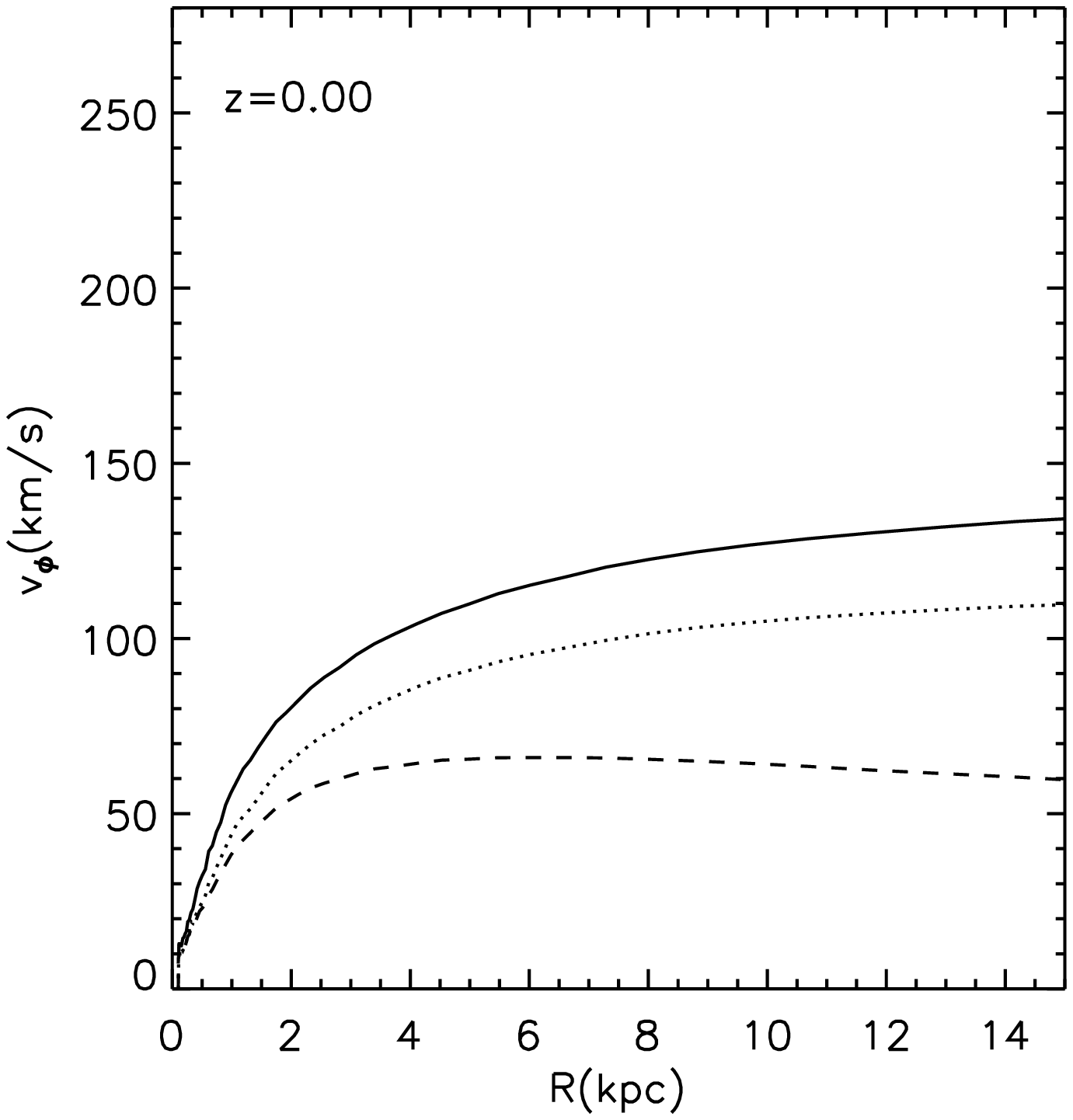}
	}\\	

	\subfigure[\MWone\hspace{0.35\gsize}]
	{
		\label{fig1i}
		\includegraphics[height=\gsize]{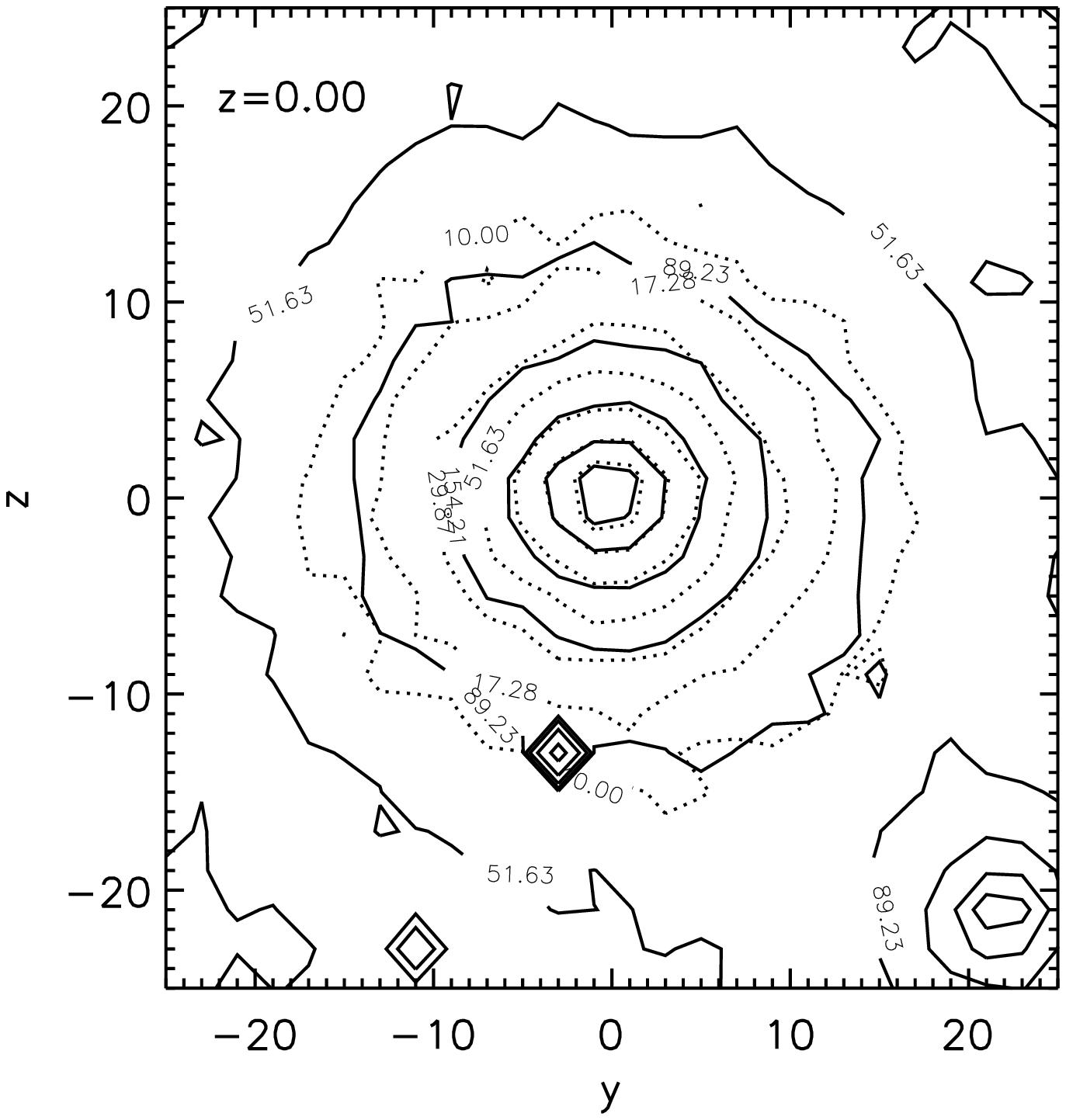}
	}
	\hspace{-0.2\gsize}
	\subfigure[\MWtwo\hspace{0.35\gsize}]
	{
		\label{fig1j}
		\includegraphics[height=\gsize]{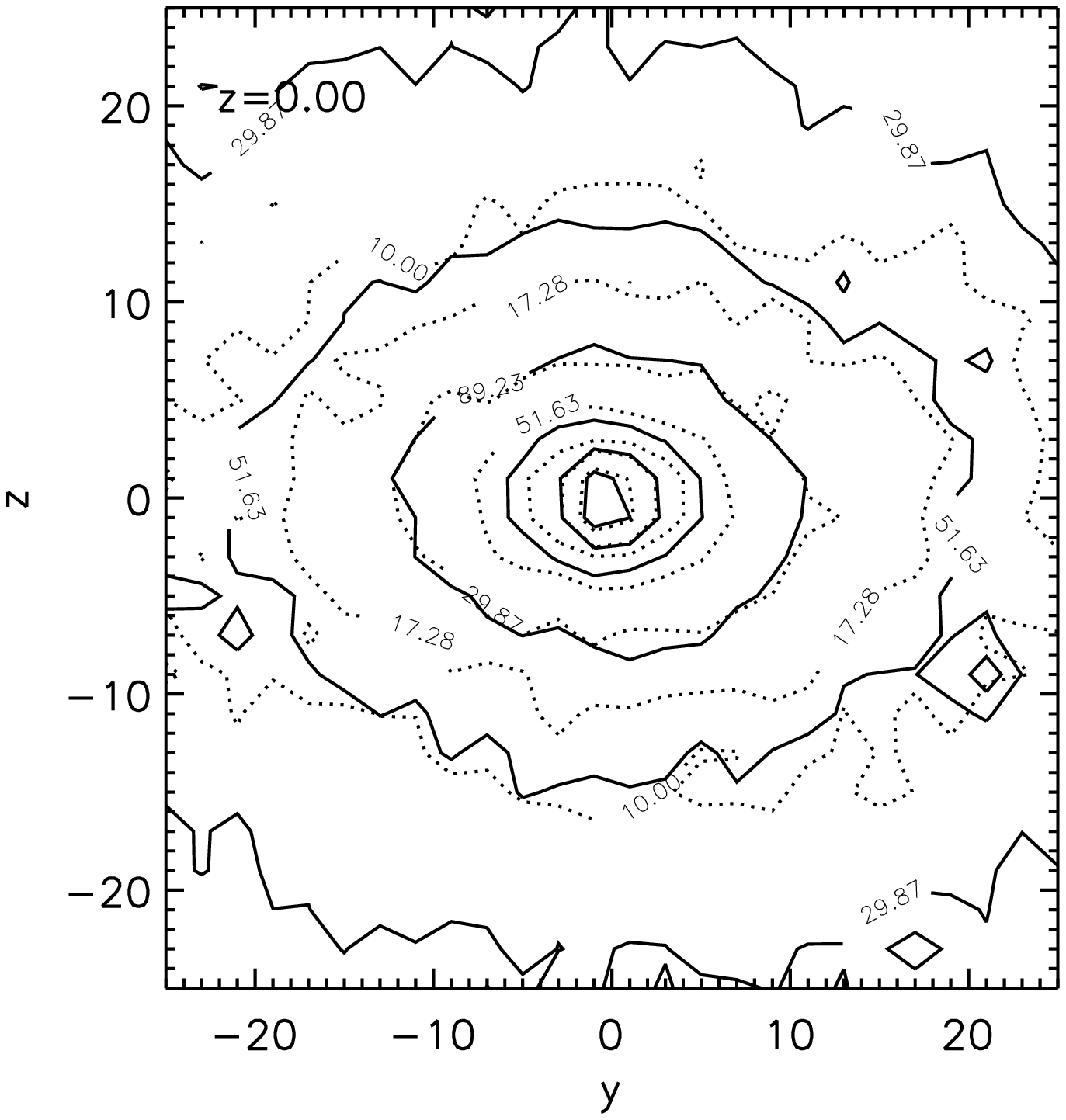}
	}
	\hspace{-0.2\gsize}
	\subfigure[\MWthree\hspace{0.35\gsize}]
	{
		\label{fig1k}
		\includegraphics[height=\gsize]{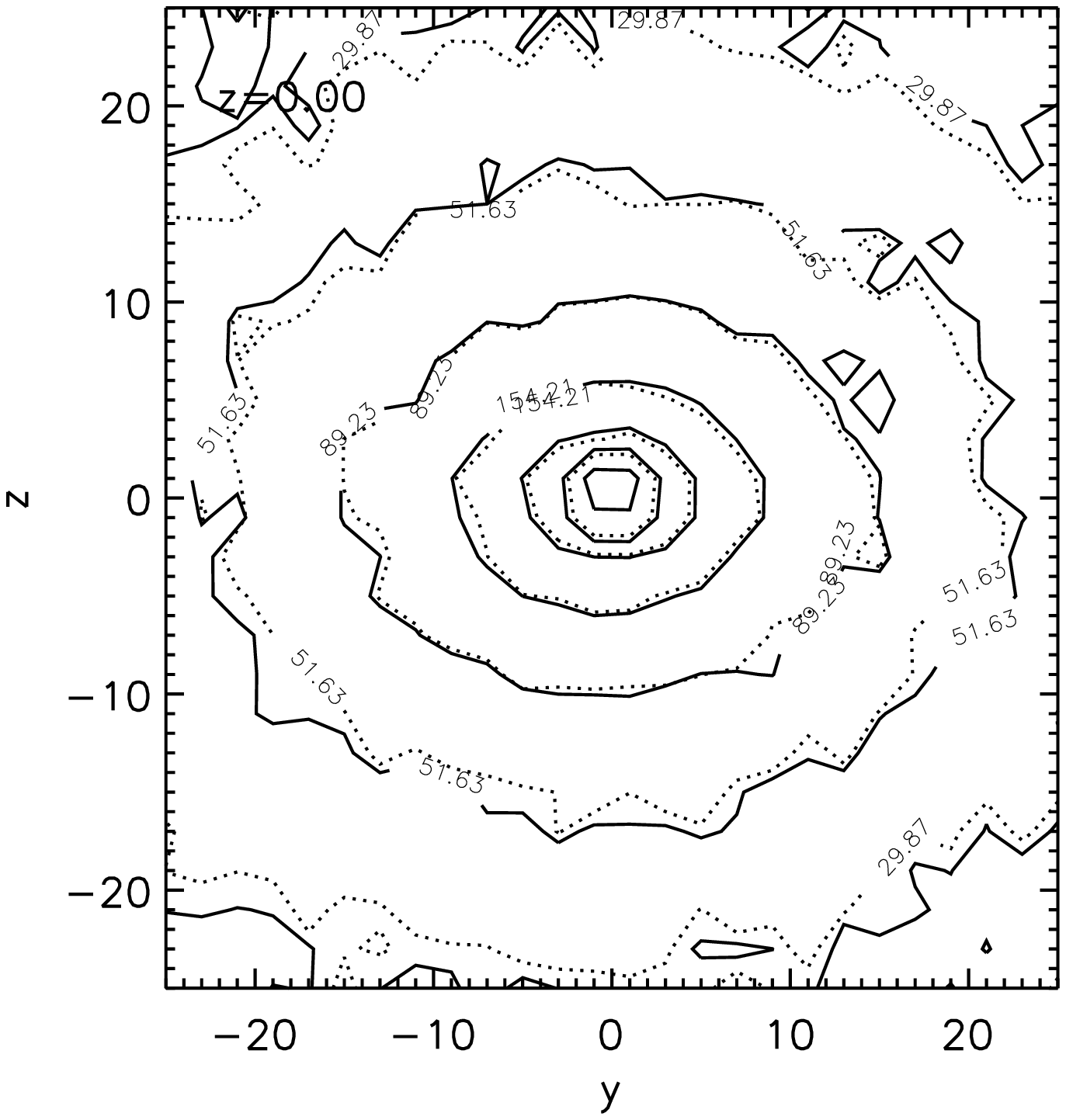}
	}
	\hspace{-0.2\gsize}
	\subfigure[\MWthreedark\hspace{0.22\gsize}]
	{
		\label{fig21l}
		\includegraphics[height=\gsize]{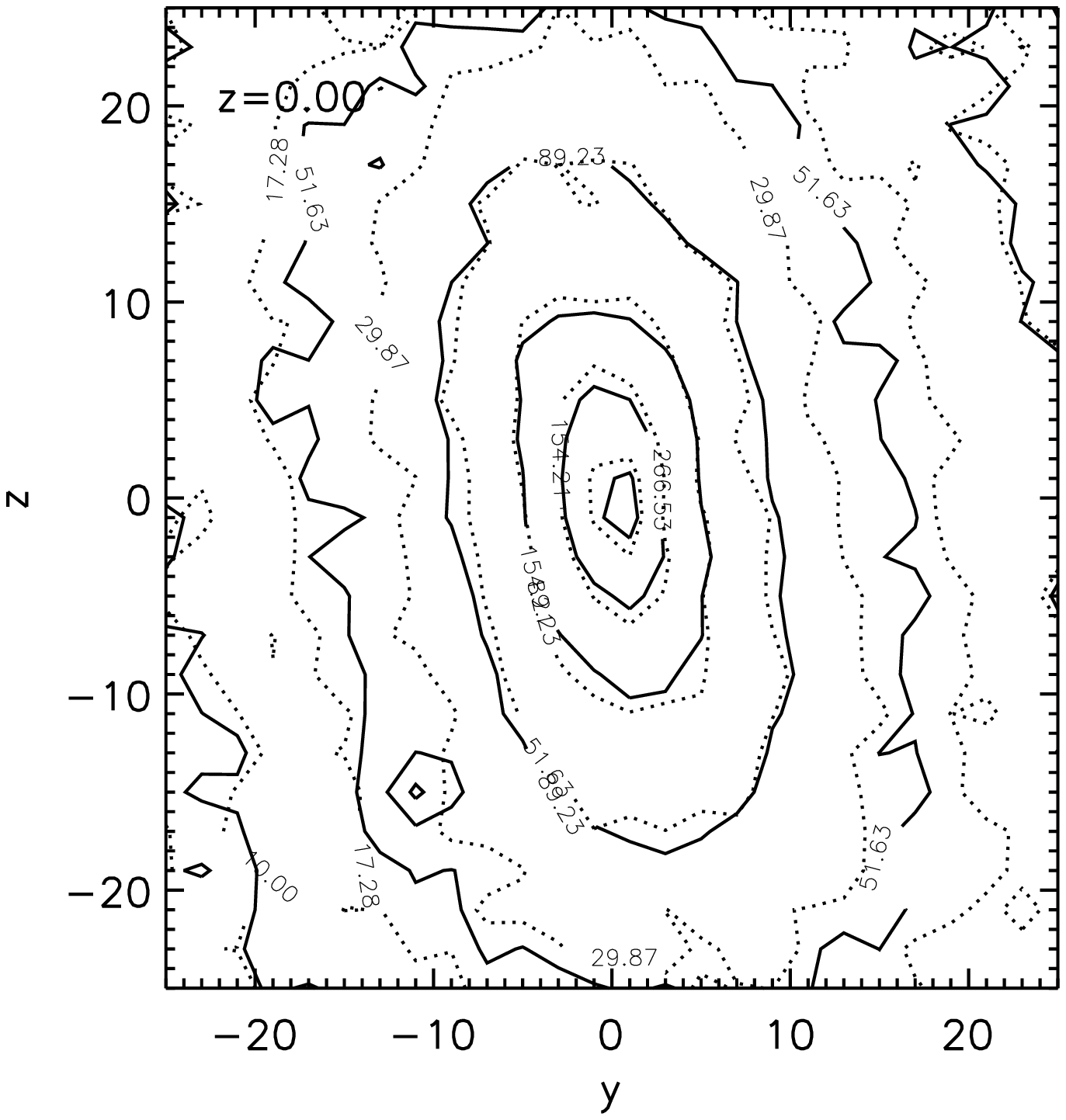}
	}\\	

	\subfigure[\MWone\hspace{0.35\gsize}]
	{
		\label{fig1u}
		\includegraphics[height=\gsize]{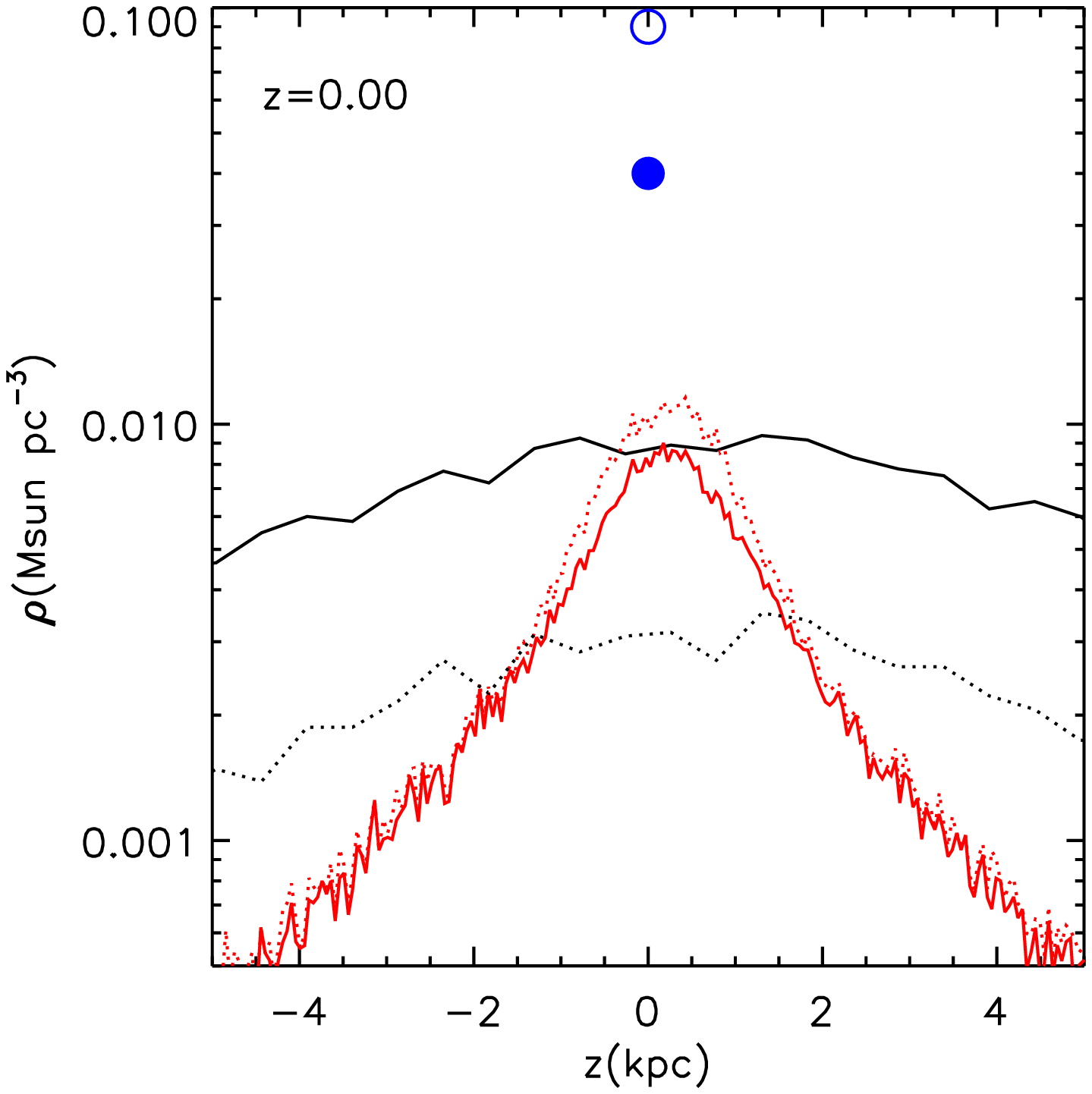}
	}
	\hspace{-0.2\gsize}
	\subfigure[\MWtwo\hspace{0.35\gsize}]
	{
		\label{fig1v}
		\includegraphics[height=\gsize]{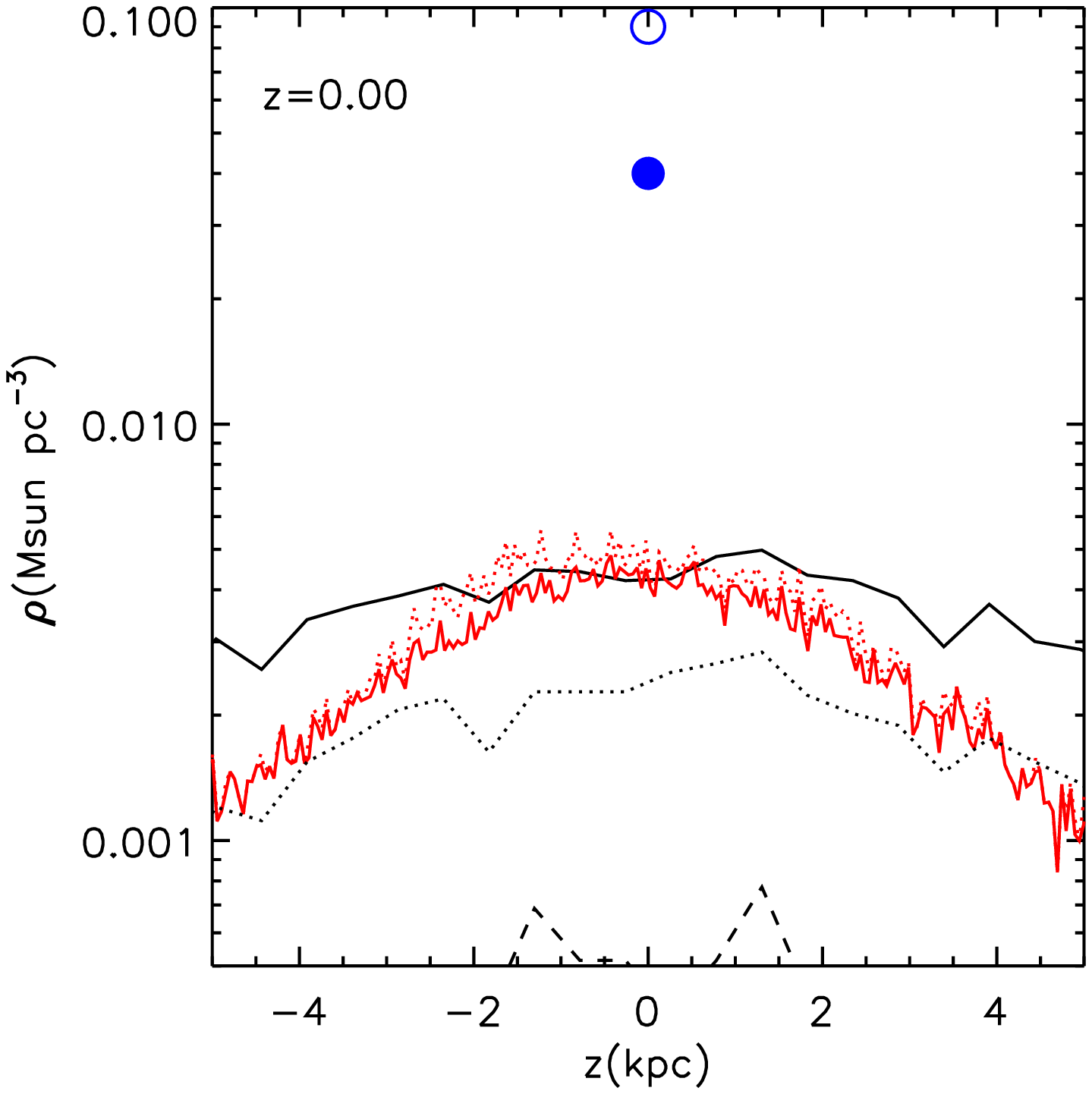}
	}
	\hspace{-0.2\gsize}
	\subfigure[\MWthree\hspace{0.35\gsize}]
	{
		\label{fig1w}
		\includegraphics[height=\gsize]{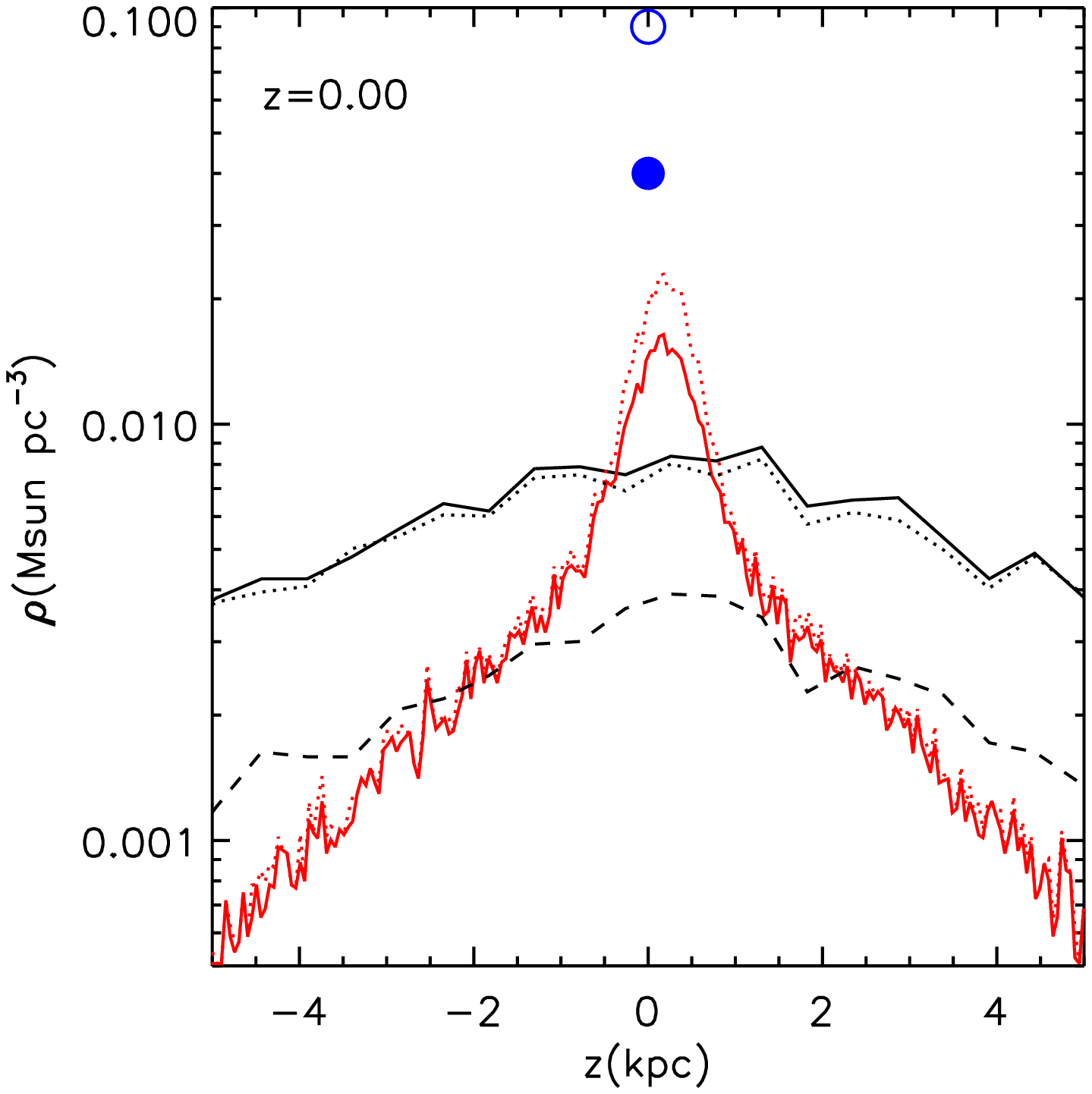}
	}
	\hspace{-0.2\gsize}
	\subfigure[\MWthreedark\hspace{0.22\gsize}]
	{
		\label{fig21x}
		\includegraphics[height=\gsize]{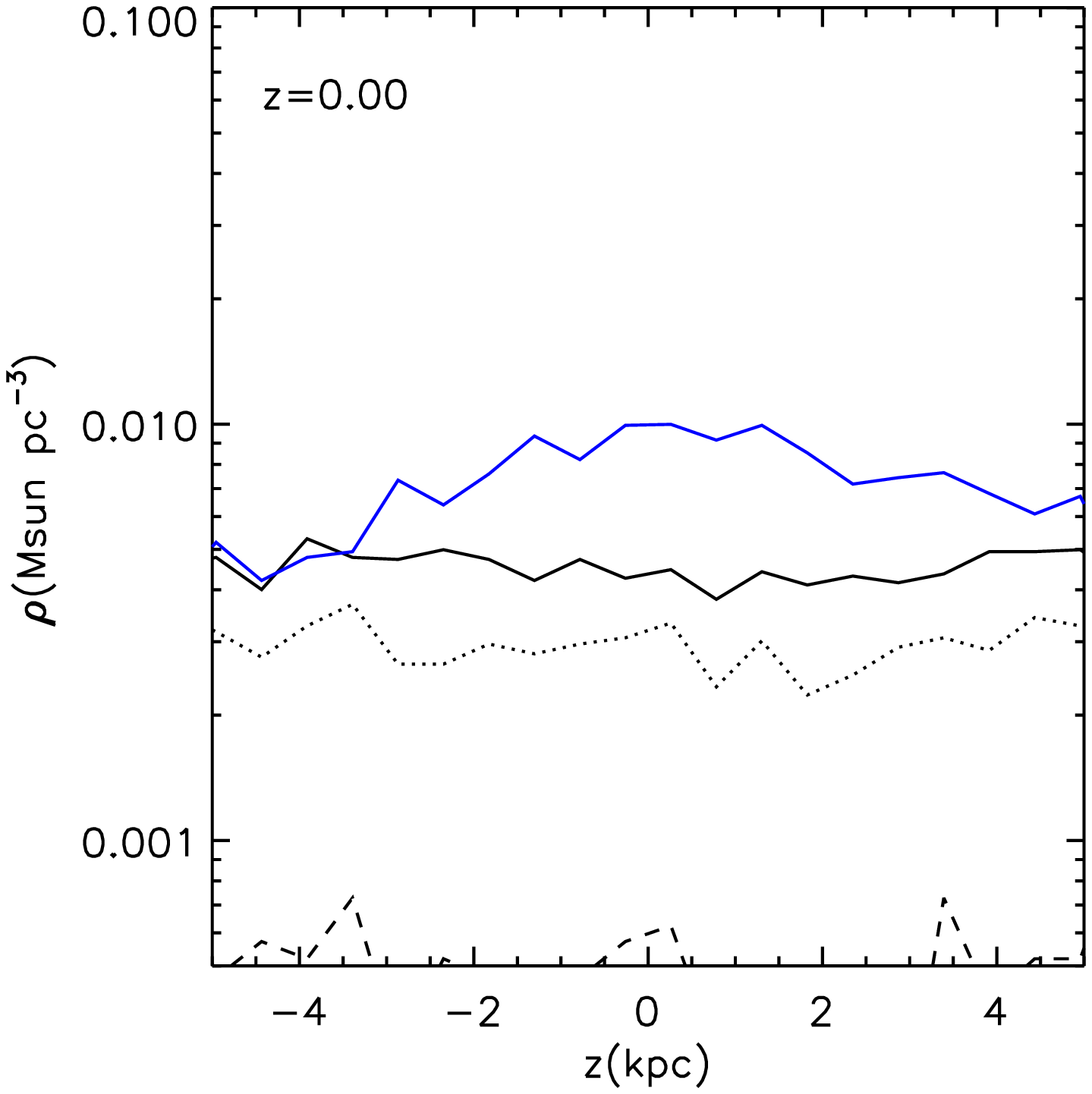}
	}\\	

\caption{{\bf (a-d)} The mean streaming velocities for the young stars (red) and gas (red dotted) in a slice $|z|<1.1$\,kpc for \MWall\ and \MWthreedark. Over-plotted are the circular velocity profiles (computed from the enclosed mass) for the dark matter (black), the sum of dark matter accreted from the four most massive satellites (black dotted), and the sum of dark matter accreted from the inner 10\% of the four most massive satellites (black dashed). {\bf (e-h)} Projected density contours for the dark matter (black) and the sum of dark matter accreted from the four most massive satellites (black dotted). {\bf (i-l)} Density of stars (red), stars+gas (red dotted), dark matter (black) and dark matter accreted from the four most massive satellites (black dotted and dashed as in (a-d)) as function of height $z$ in a slice $7<R<8$\,kpc for \MWall\ and \MWthreedark. For \MWall, we overplot the observed star (solid blue circle) and star+gas (open blue circle) stellar density in the solar neighbourhood for the Milky Way (taken from \bcite{2000MNRAS.313..209H}). For \MWthreedark, we show the profile for the alignment given in panel (h) (black), and one at 90 degrees to this (blue).}
\label{fig:cosmo2}
\vspace{-3mm}
\end{figure*}
\end{center}

\begin{center}
\begin{figure*}
	\setlength{\gsize}{0.27\textwidth}
	\setlength{\subfigcapskip}{-1.05\gsize}

	\subfigure[\MWone\hspace{0.35\gsize}]
	{
		\label{fig3a}
		\includegraphics[height=\gsize]{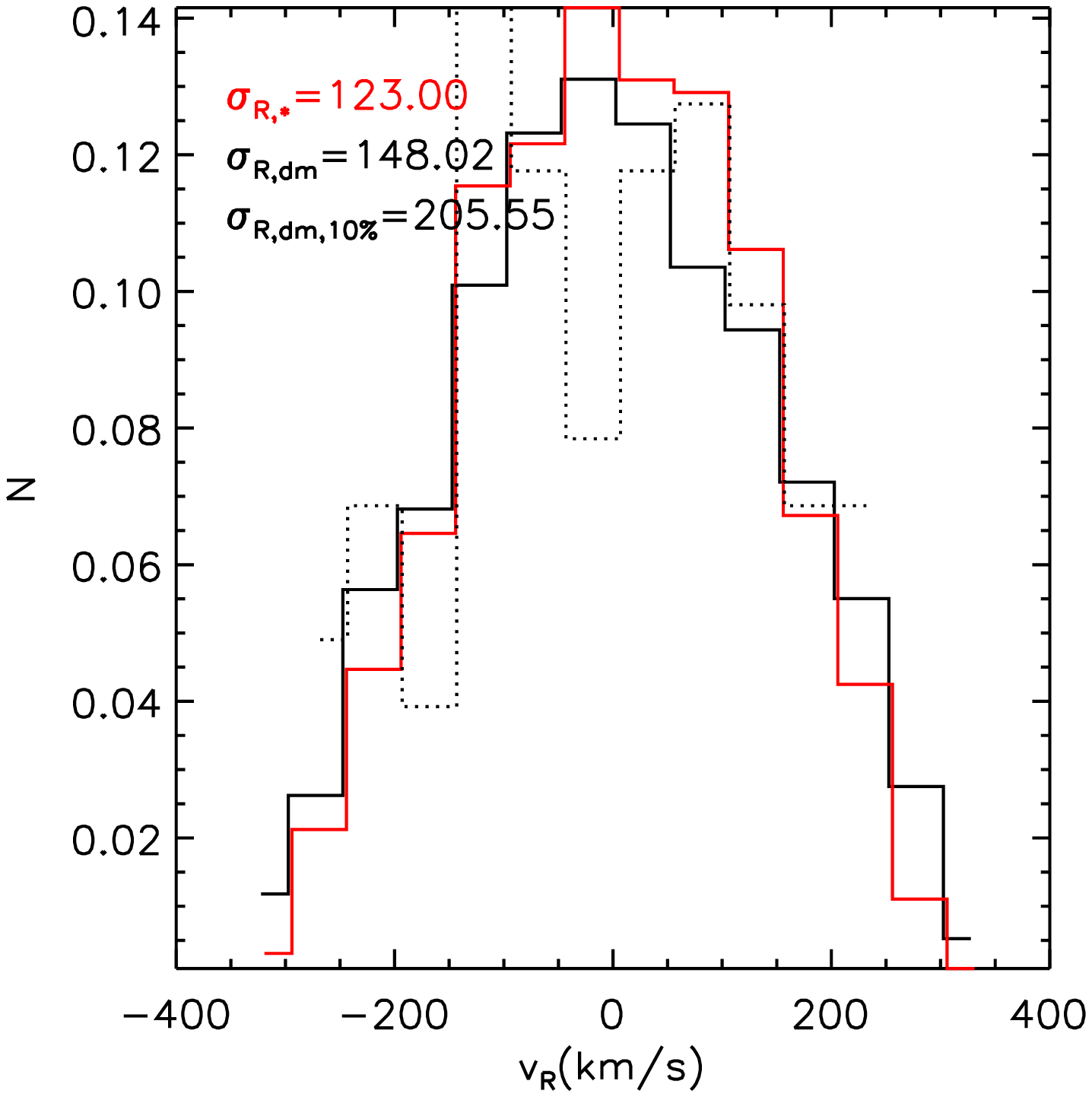}
	}
	\hspace{-0.2\gsize}
	\subfigure[\MWone\hspace{0.35\gsize}]
	{
		\label{fig3b}
		\includegraphics[height=\gsize]{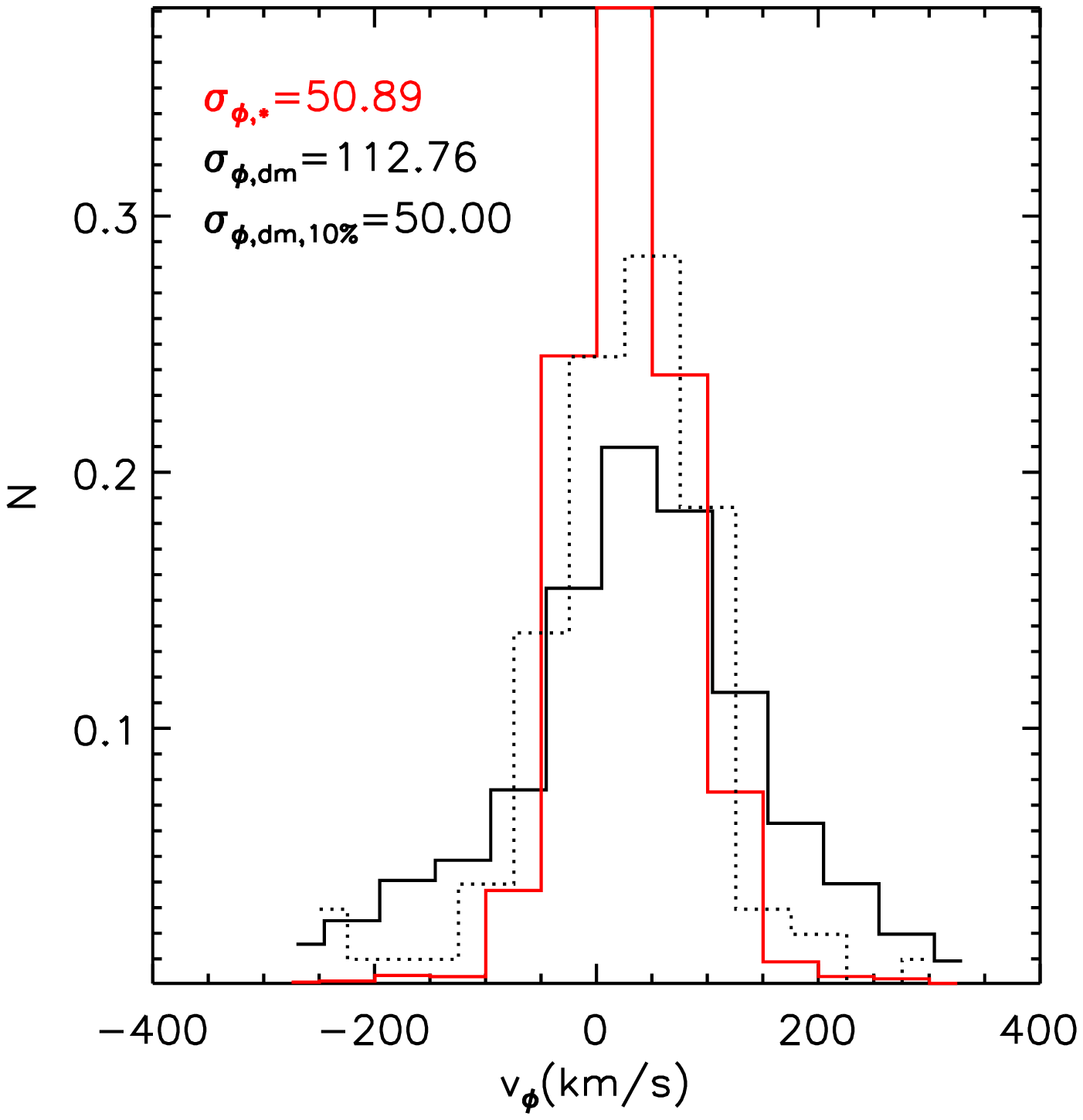}
	}
	\hspace{-0.2\gsize}
	\subfigure[\MWone\hspace{0.35\gsize}]
	{
		\label{fig3c}
		\includegraphics[height=\gsize]{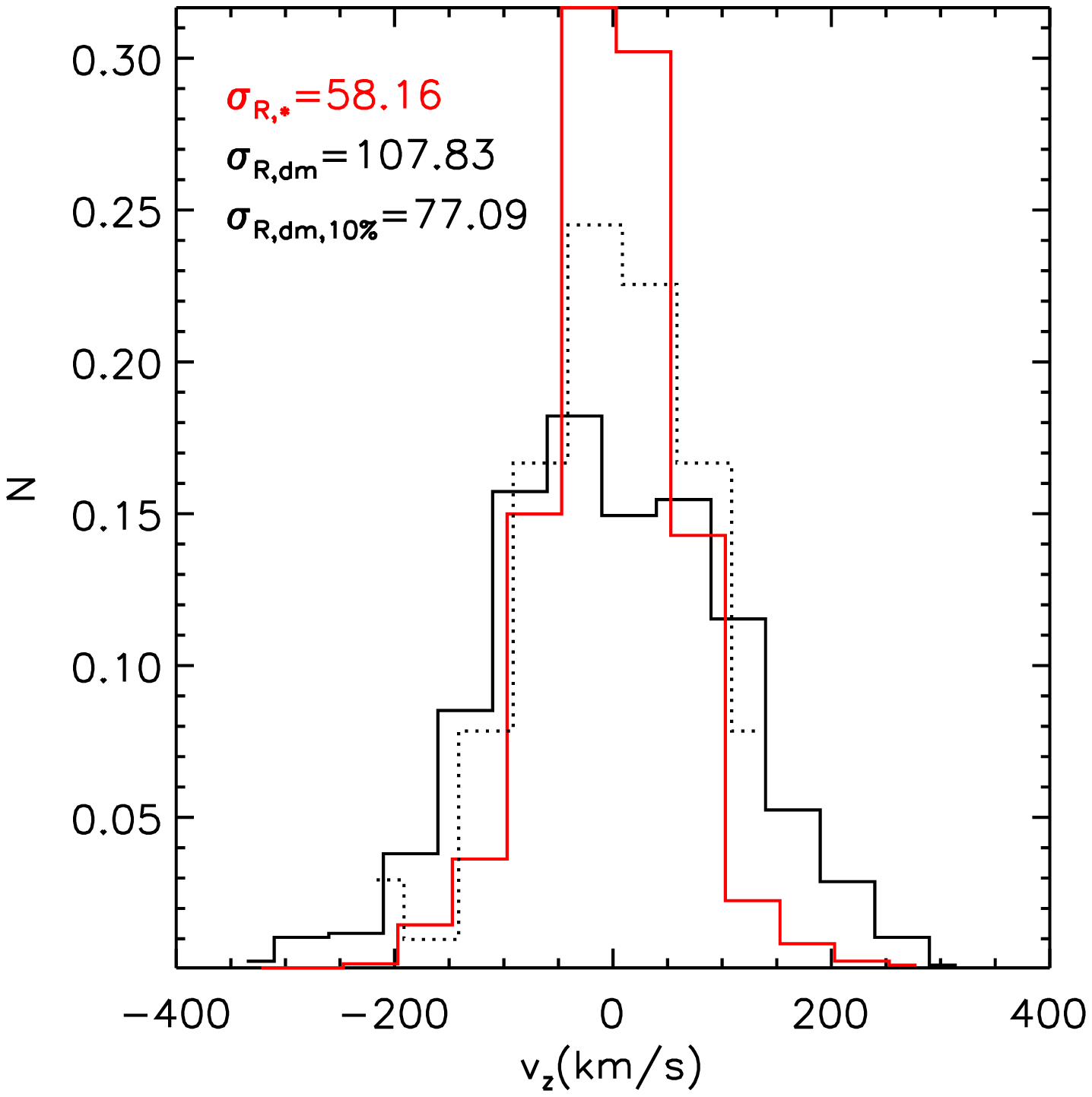}
	}\\

	\subfigure[\MWtwo\hspace{0.35\gsize}]
	{
		\label{fig3d}
		\includegraphics[height=\gsize]{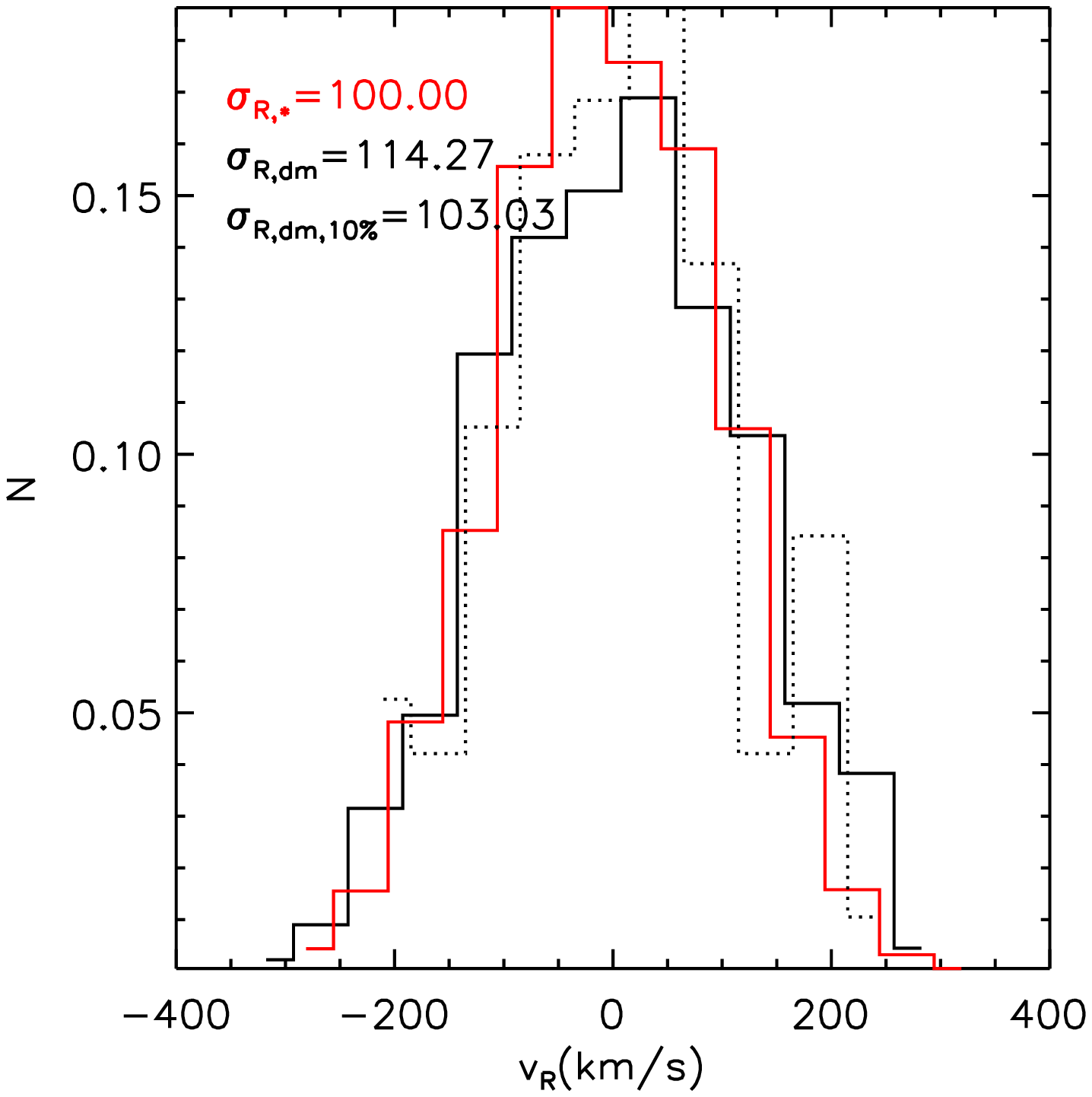}
	}
	\hspace{-0.2\gsize}
	\subfigure[\MWtwo\hspace{0.35\gsize}]
	{
		\label{fig3e}
		\includegraphics[height=\gsize]{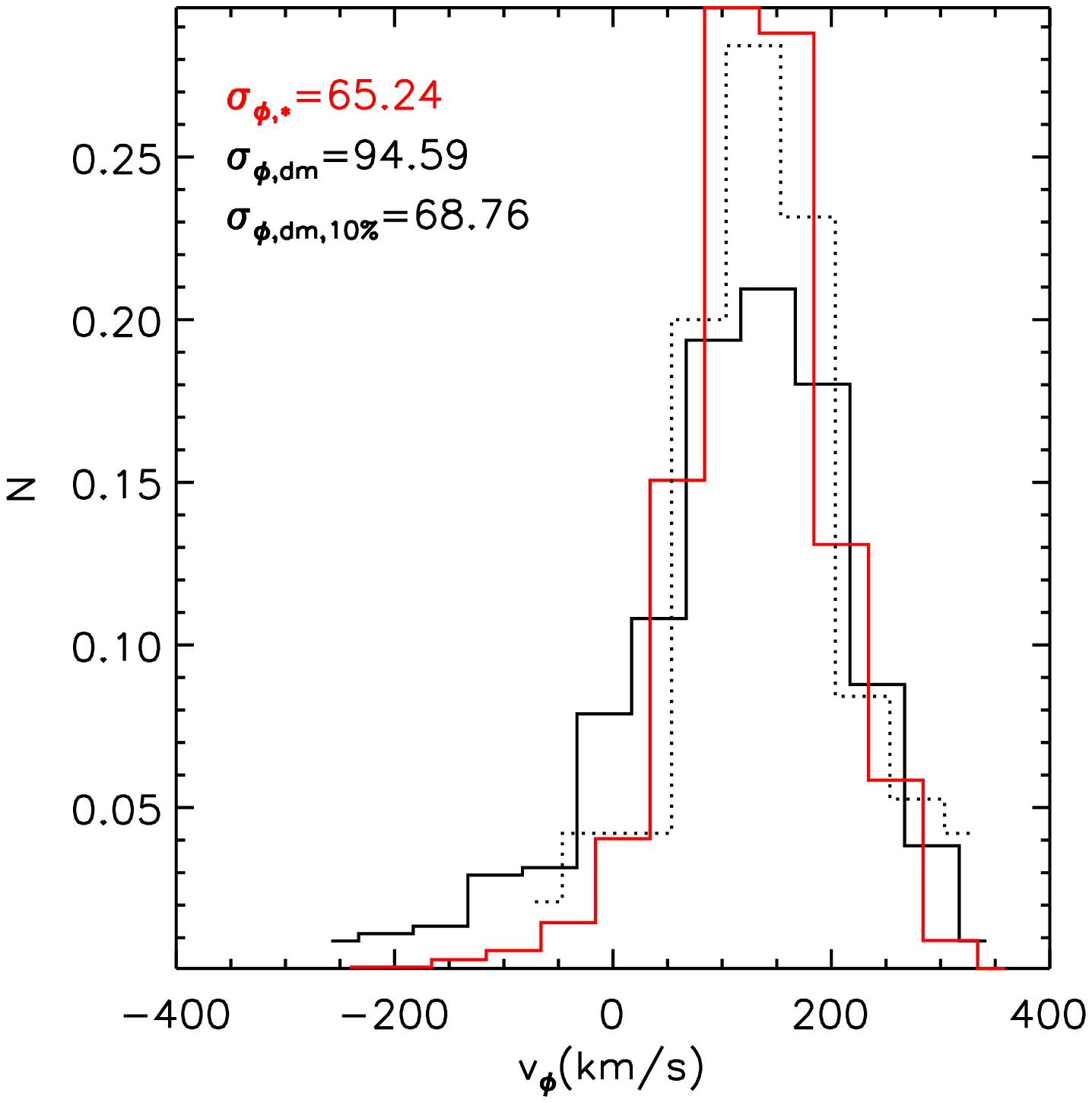}
	}
	\hspace{-0.2\gsize}
	\subfigure[\MWtwo\hspace{0.35\gsize}]
	{
		\label{fig3f}
		\includegraphics[height=\gsize]{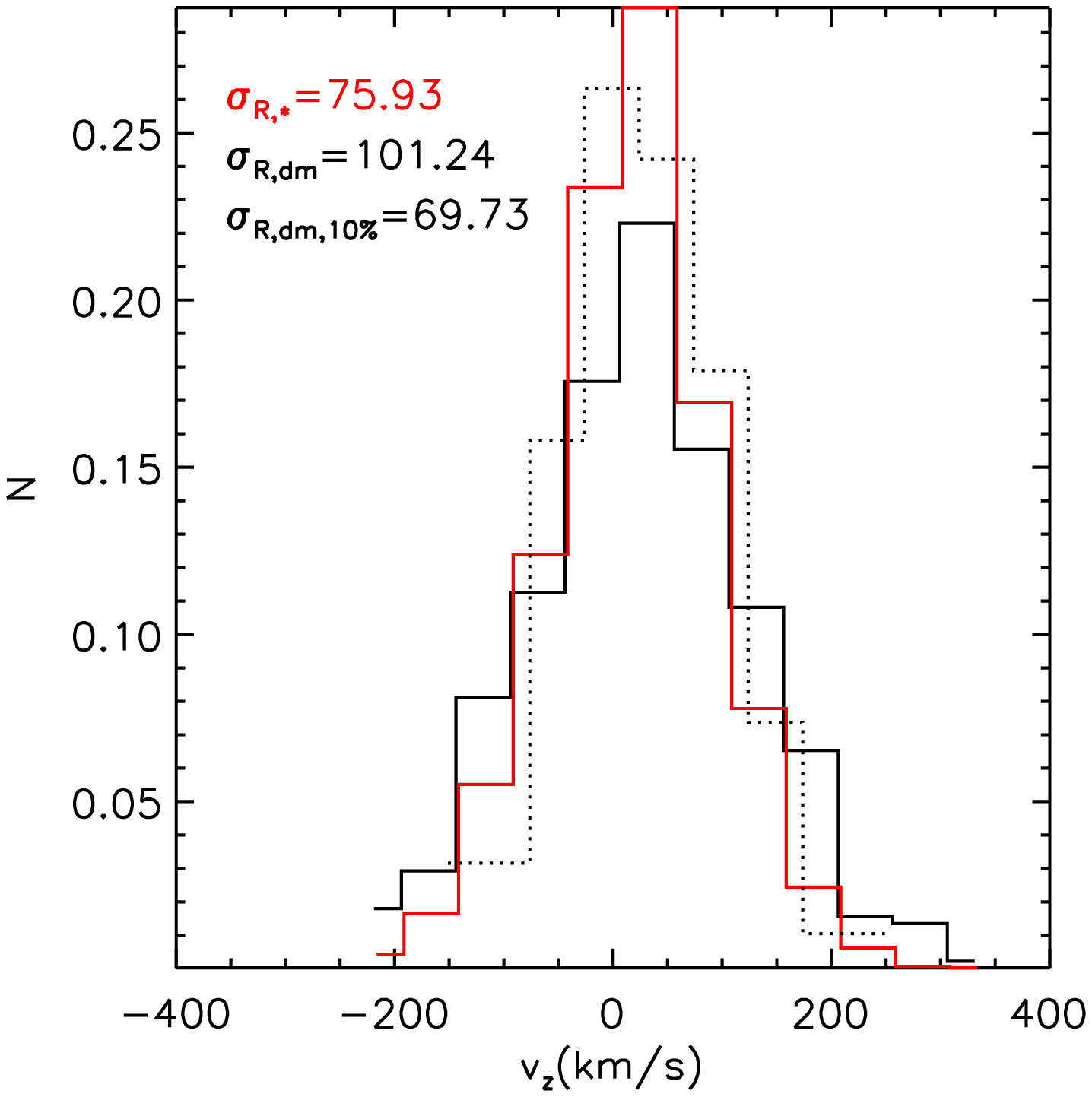}
	}\\

	\subfigure[\MWthree\hspace{0.35\gsize}]
	{
		\label{fig3g}
		\includegraphics[height=\gsize]{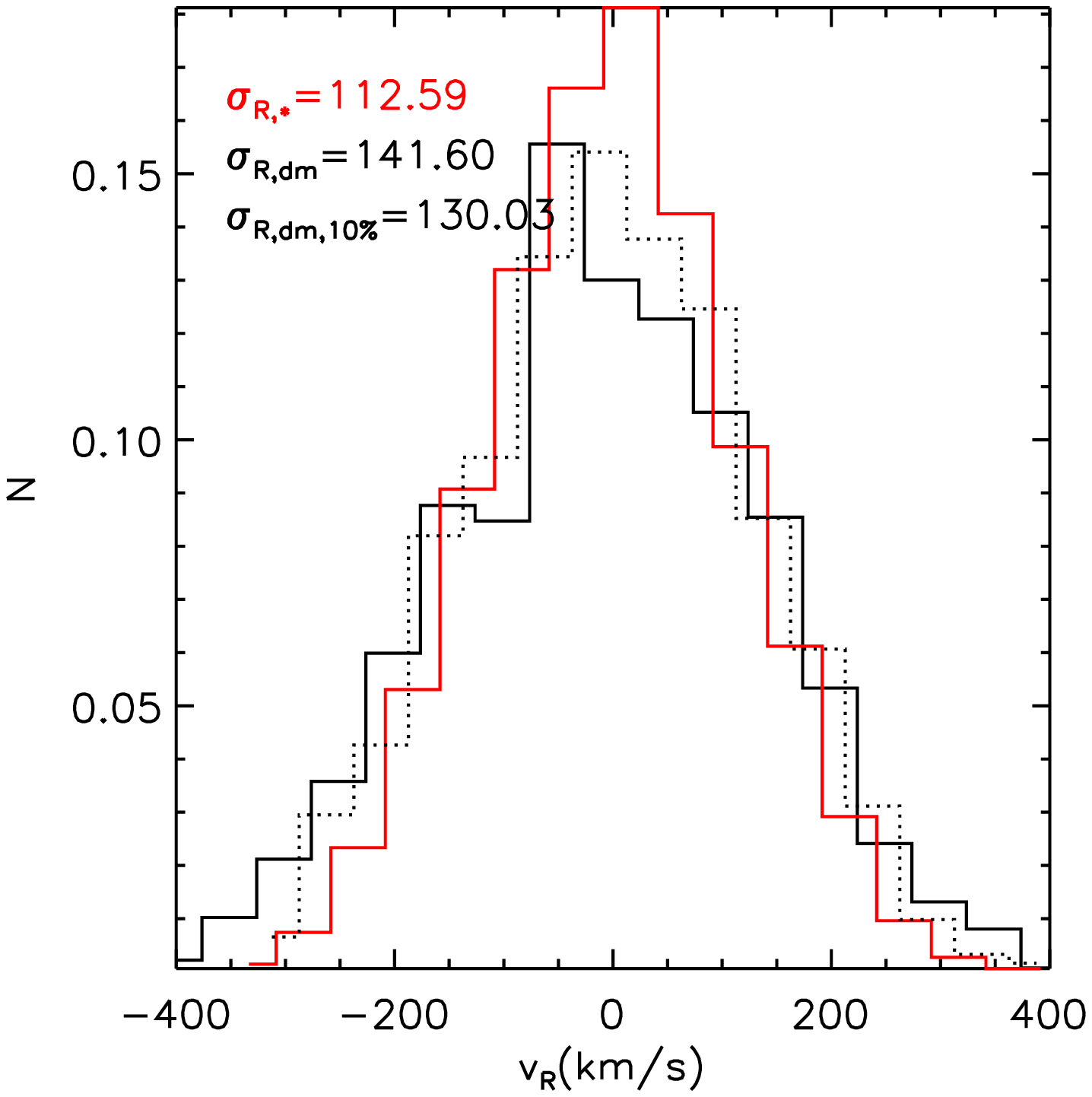}
	}
	\hspace{-0.2\gsize}
	\subfigure[\MWthree\hspace{0.35\gsize}]
	{
		\label{fig3h}
		\includegraphics[height=\gsize]{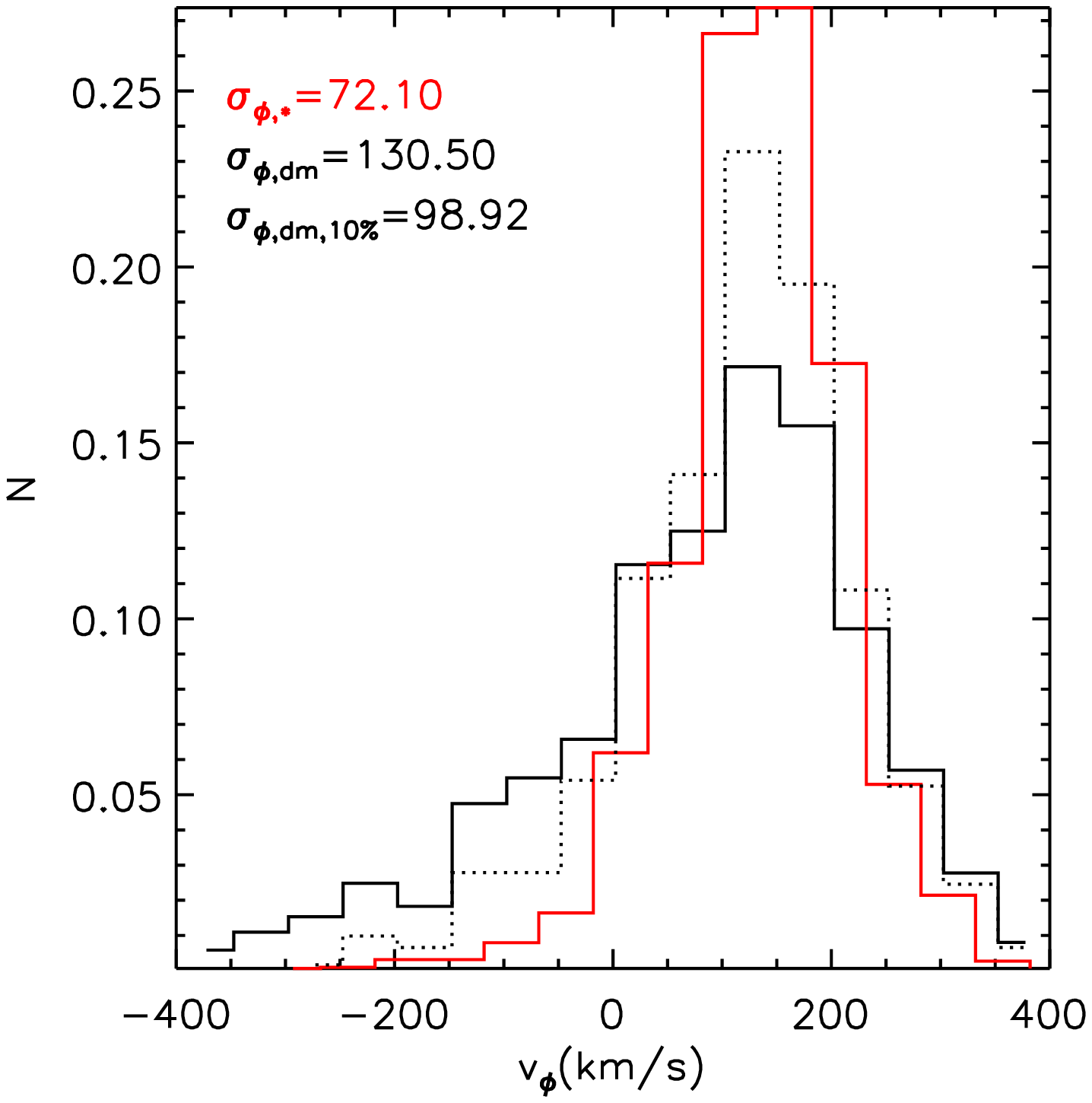}
	}
	\hspace{-0.2\gsize}
	\subfigure[\MWthree\hspace{0.35\gsize}]
	{
		\label{fig3i}
		\includegraphics[height=\gsize]{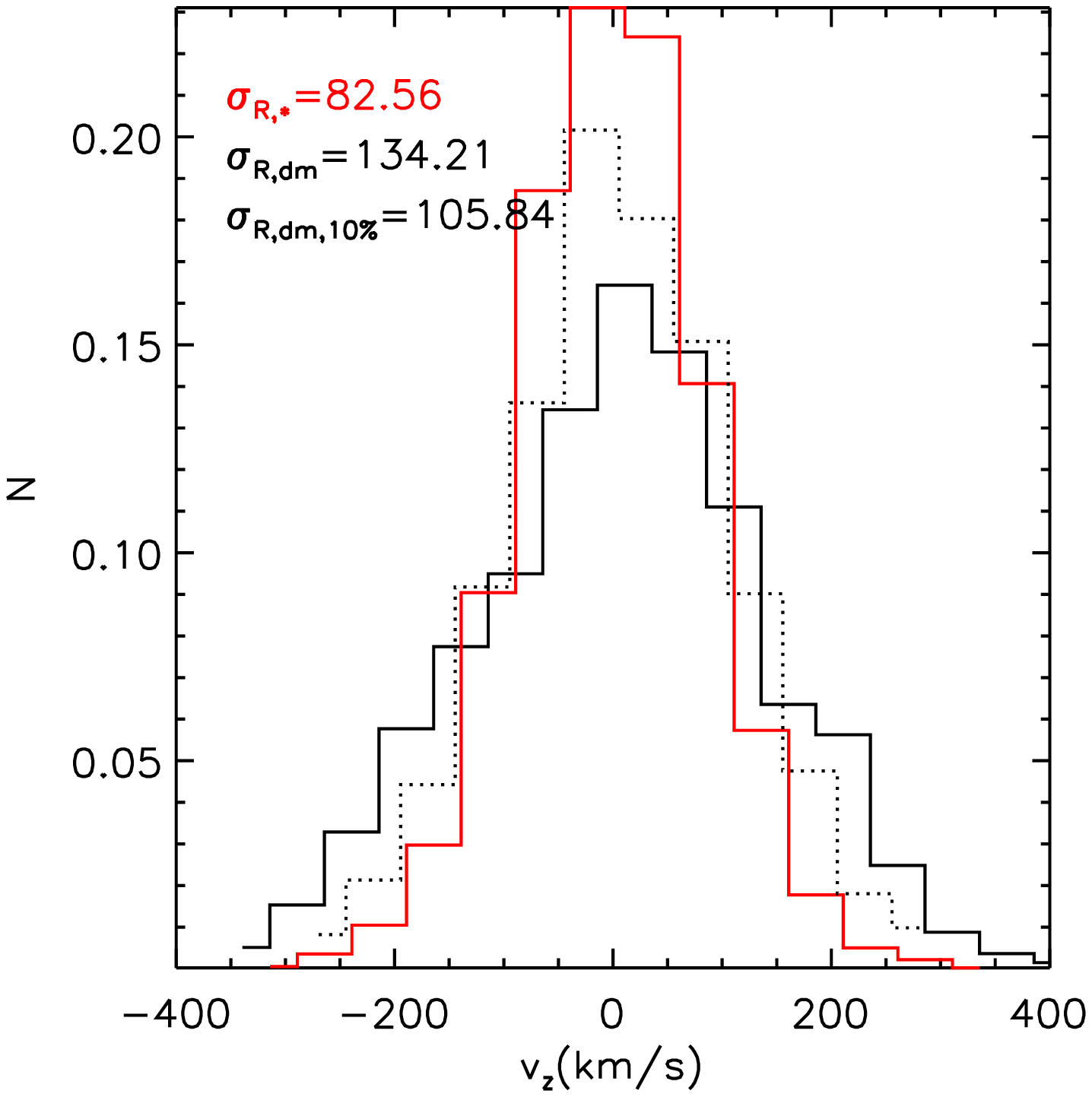}
	}\\
		
\caption{{\bf (a-d)} A comparison of the accreted star (red) and accreted dark matter (black) velocity distributions $v_R$ (left), $v_\phi$ (middle) and $v_z$ (right) at the solar neighbourhood ($7<R<8$\,kpc; $|z|<2.1$\,kpc), for \MWone, \MWtwo\ and \MWthree. We plot the sum of material accreted from the four most massive disrupted satellites. Also plotted is the dark matter coming from the inner 10\% of these satellites (dotted lines). The dispersions are given in the top left corner of each plot in km/s.}
\label{fig:cosmo3}
\vspace{-3mm}
\end{figure*}
\end{center}

\section{Results}\label{sec:results}

\subsection{Forming a disc of dark matter}\label{sec:darkdisc}

Figure \ref{fig:precosmo}(a-c) shows the distribution of rotational velocities $v_\phi$ at the solar neighbourhood ($7<R<8$\,kpc; $|z|<2.1$\,kpc) for the stars (red) and dark matter (black) in \MWone, \MWtwo\ and \MWthree. Overplotted in panel (c) is the distribution for \MWthreedark, modelled with dark matter alone (black dotted). Our results are not sensitive to the position of the `solar neighbourhood' slice (a choice of $5.5<R<6.5$\,kpc gave near-identical results). With the exception of \MWthreedark, all of the dark matter distributions are skewed towards the stars. Figure \ref{fig:cosmo}(a-d) shows a decomposition of these distributions. All three baryonic simulations require a double Gaussian fit to their dark matter $v_\phi$ distribution (smooth black and black dotted lines). The second Gaussian rotates rapidly and matches well the material accreted from the sum of the four most massive satellites (black dotted histogram), though our fit did not require this. (Nor did our fit require either Gaussian to have mean rotational velocity $\overline{v_\phi}=0$.) It is this accreted, rapidly rotating material that we call the `dark disc'.

Figure \ref{fig:cosmo}(a-d) shows how the material accreted from the four most massive merging satellites\footnote{Considered over their whole lifetime.} (green, blue, magenta and cyan) contributes to the dark disc for \MWone, \MWtwo\ and \MWthree. Figure \ref{fig:cosmo}(e-h) shows the decay in radius as a function of redshift $z$ of these satellites; and Figure \ref{fig:cosmo}(i-l) shows the decay in angle $\theta$ to the host galaxies' disc as a function of redshift $z$. 

The mass and rotation speed of the dark disc increase in the simulations with more late mergers. \MWone\ has no significant mergers after redshift $z=2$ and has a less significant dark disc, with rotation lag with respect to the stars (red lines in Figure \ref{fig:precosmo}) of $\sim150$\,km/s, and dark disc to halo density ratio of $\rhodd/\rhoh = 0.23$ (obtained from the double Gaussian fit). \MWtwo\ and \MWthree\ both have extreme dark discs with $\rhodd/\rhoh > 1$ and rotation lag with respect to the stars of $\simlt 60$\,km/s; they both have massive mergers at redshift $z<1$. 

Figure \ref{fig:cosmo}(i-j) demonstrates that disc plane dragging is responsible for the formation of the dark disc. In \MWone, the green satellite is dragged towards the disc plane, the magenta and cyan satellites start out close to the disc plane, and the blue satellite merges at high inclination angle. Figure \ref{fig:cosmo}(a) shows the contribution to the dark disc owing to each of these satellites. The magenta satellite contributes the most, being both low-inclination and massive, then the green. The cyan satellite is of too low mass to contribute significantly, while the blue satellite contributes little rotating material because of its high inclination. These results confirm our expectations from isolated disc-satellite merger simulations \citep{2008MNRAS.389.1041R}. Similar results can be seen for the four most massive mergers in \MWtwo. Although initially on high inclination orbits, the magenta and cyan satellites complete enough peri-centre passages to be dragged down into the disc plane and contribute significantly to the dark disc; the blue and green satellites also contribute in equal measure, though somewhat less than the magenta and cyan satellites owing to their higher final inclinations. \MWthree\ appears to present a similar picture. The green satellite is a near $\sim$1:1 merger that starts out near the disc plane and contributes nearly all of the dark disc. However, mergers of this mass ratio define the post-merger plane of the disc. They will lead to highly rotating, albeit hot, dark matter discs, whatever their initial inclination. Yet there is no dark disc in \MWthreedark\ that has the same $\sim$1:1 merger, but no baryonic material. This suggests that a second mechanism, extra to disc-plane dragging, is important for the formation of dark discs. We discuss this below in \S\ref{sec:darkdisckeep}.

It is interesting that in all three galaxies, none of the four most massive satellites contribute significant retrograde material. We will investigate this further in future work, but note here that retrograde mergers are suppressed both because of reduced dynamical friction (\bcite{1986ApJ...309..472Q}; \bcite{2008MNRAS.389.1041R}), and reduced tidal forces \citep{2006MNRAS.366..429R}. 

\subsection{Maintaining the dark disc: the importance of halo shape}\label{sec:darkdisckeep}

Figure \ref{fig:cosmo2}(e-h) shows projected density contours for the total dark matter (black) and the dark matter accreted from the four most massive satellites (black dotted), in \MWall, and \MWthreedark. All of the simulations that include the baryons produce near-spherical, slightly oblate, final dark matter density distributions that have symmetry axes parallel to the stellar disc. This agrees well with previous numerical results in the literature (\bcite{2004ApJ...611L..73K}; \bcite{2007arXiv0707.0737D}). Indeed, observations of the Sagittarius stream of stars suggest that the Milky Way has a near-spherical halo with axis ratio $a/c \simgt 0.9$ within $\sim 50$\,kpc (\bcite{2005ApJ...619..800J}; \bcite{2006ApJ...651..167F}). Of our three simulated galaxies, only the most flattened case, \MWtwo\ with an axis ratio $a/c \sim 0.77$ within $\sim 50$\,kpc, is inconsistent with the Milky Way. By contrast, without the baryons, the dark matter halo in \MWthreedark\ is highly triaxial (Figure \ref{fig:cosmo2}(h)). Indeed, the typical dark matter halo in $\Lambda$CDM -- in simulations that model the dark matter alone -- is triaxial, prolate, and inconsistent with observations of the Milky Way \citep{2007MNRAS.378...55M}.

As noted above, both \MWthree\ and \MWthreedark\ have $\sim$1:1 mergers at redshift $z\sim1$. In \MWthree\ this merger produces a dark disc, but in \MWthreedark\ it does not. Since mergers of this mass ratio define the post-merger plane of the disc, the difference cannot be due to disc plane dragging. Instead the difference is due to the halo shapes. In a static oblate potential like that in \MWthree, particles conserve the z-component of their angular momentum vector. Any sense of rotation established in the merger will be preserved. By contrast, in a static {\it triaxial} potential like that in \MWthreedark, particles moving on regular orbits do not explicitly conserve any component of their angular momentum vector. Their orbit planes precess and any sense of rotation established in the merger is rapidly lost. As a result, the final velocity distribution at the solar neighbourhood in \MWthreedark\ is a Gaussian. This demonstrates that in addition to disc plane dragging, the near-spherical halo that results once the baryons are included is a vital ingredient in the formation and survival of a dark matter disc.

\subsection{The mass and spatial distribution of the dark disc}\label{sec:darkdiscspat}

The precise properties of the dark disc depend on our choice of decomposition. Here, we define the dark disc as the material that originates from the innermost 10\% of the four most massive merging satellites. This gives a reasonable match to the highly rotating Gaussian component in our double Gaussian fit to the local dark matter velocity distribution (see \S\ref{sec:observe}). 

Figure \ref{fig:cosmo2}(a-d) shows the mean streaming velocities ($\overline{v_\phi}$ averaged in bins $\Delta R = 1$\,kpc) for the stars (red) and gas (red dotted) in a slice $|z|<1.1$\,kpc for \MWall\ and \MWthreedark. Over-plotted are the circular velocity profiles (computed from the enclosed mass) for the dark matter (black), the sum of dark matter accreted from the four most massive satellites (black dotted), and the sum of dark matter accreted from the inner 10\% of the four most massive satellites (black dashed). Using our above decomposition, the dark disc contributes at most half of the rotation curve at the solar neighbourhood (in \MWthree). Since the rotation curve goes as the square root of the mass, this is a quarter of the mass. The other extreme is \MWone\ where the dark disc contributes a quarter of the rotation curve at the solar neighbourhood, which is $1/16$ of the mass. 

Figure \ref{fig:cosmo2}(i-l) shows the density of stars (red), dark matter (black), dark matter accreted from the four most massive satellites (black dotted), and dark matter accreted from the inner 10\% of the four most massive satellites (black dashed) as a function of height $z$ in a slice $7<R<8$\,kpc for \MWall\ and \MWthreedark. In \MWone, \MWtwo\ and \MWthree\ the density profile is peaked, indicating a disc-like structure. In \MWthreedark, the profile depends on our choice of alignment. In its eigenframe, $\rho(z)$ is flat, as expected for a more spheroidal dark matter distribution. In a frame rotated 90 degrees to this, however, the distribution is peaked similarly to \MWthree\ (blue line; Figure \ref{fig:cosmo2}(l)). However, there is no measurable rotation in the dark matter $v_\phi$ distribution for \MWthreedark\ for any alignment. The peaky $\rho(z)$ distribution in \MWthreedark\ is the result of a flattened halo, not a dark disc. For our above decomposition, the dark disc in \MWone, \MWtwo\ and \MWthree\ is a structure that rotates in the same sense as the stellar disc and is flattened in the same plane as the stellar disc -- hence the name `dark disc'. 

\subsection{Observing the dark disc}\label{sec:observe}

Figure \ref{fig:cosmo3} shows a comparison of the star (red) and dark matter (black) velocity distributions at the solar neighbourhood ($7<R<8$\,kpc; $|z|<2.1$\,kpc) of material accreted from the four most massive disrupted satellites in \MWone, \MWtwo\ and \MWthree. Also plotted is the dark matter coming from the inner 10\% of these satellites (dotted line). The accreted stars comprise 5, 40 and 30\% of the stars in the solar neighbourhood for \MWone, \MWtwo\ and \MWthree, respectively (see \bcite{brookinprep} for a study on the decomposition of these into bulge, disc and halo stars). The accreted stars are colder than the full accreted dark matter distributions, but closer to the inner 10\% dark matter distribution. The best-fit parameters for the dark disc (see Figure \ref{fig:cosmo}(a-d)) lie somewhere in-between the stellar distributions and the inner 10\% dark matter distributions. If future surveys of our Galaxy can disentangle accreted stars in the Milky Way thick disc from those that formed in-situ, then we will be able to infer, through numerical modelling, the velocity distribution function of the dark disc from these stars. We will investigate this in more detail in future work. 

\section{Discussion: the importance of baryons}\label{sec:discussion} 

The simulations \MWall\ and \MWthreedark\ demonstrate the importance of including the baryons in simulations of the local dark matter distribution. Once the stars and gas are included, the dark matter halos become significantly rounder, while the stellar and gas disc biases the accretion of massive satellites, dragging them towards the disc plane. Both effects aid the formation of a dark disc and lead to solar neighbourhood velocity distributions that are anisotropic, and better fit in $v_\phi$ by a double Gaussian than a single Gaussian. 

It is not clear which of \MWone, \MWtwo\ or \MWthree\ provides the best match to our Galaxy. The Milky Way has evidence for several mergers since $z\sim 1$. The Sagittarius dwarf recently fell in on a polar orbit (\bcite{1994Natur.370..194I}; \bcite{2003ApJ...599.1082M}), there is evidence for a merger in the plane of the disc \citep{2007MNRAS.376..939C}, and another $\simlt 5$\,Gyrs ago from the chemistry and kinematics of stars in the solar neighbourhood \citep{2008arXiv0811.1777F}, while the Milky Way thick disc itself is evidence for a $\sim$1:10 merger at $z\sim1$ (\bcite{2008MNRAS.389.1041R}; \bcite{2007arXiv0708.1949K}). \MWone\ had no significant mergers after $z\sim2$ and is likely over-quiescent as compared to our Galaxy. By contrast, both \MWtwo\ and \MWthree\ appear to have merger histories that are too extreme as compared with the Milky Way. This results in large fractions ($30-40$\%) of accreted stars at the solar neighbourhood that are inconsistent with observations of the Milky Way. (Assuming all stars in the Milky Way thick disc are accreted gives an upper bound of 12\% for our Galaxy \citep{2008ApJ...673..864J}.)

However, differences between the simulated galaxies and the Milky Way could also be a result of numerical limitations. All three state-of-the-art simulations have a force softening of $\sim 300$\,pc, which is larger than the measured scale height of the Milky Way thin stellar disc \citep{2008ApJ...673..864J}; spurious angular momentum loss occurring in underesolved progenitors may have still affected the final mass distribution even at the high mass resolution employed here \citep{2008arXiv0801.3845M}; the sub-grid physics scheme is a source of systematic error \citep{2008arXiv0801.3845M}; and the smoothed particle hydrodynamics technique employed does not correctly resolve fluid instabilities and mixing\footnote{This is not likely to be the main source of error, however, since instabilities mostly affect the interstellar medium that is in any case poorly resolved.}
 \citep{2006astro.ph.10051A}. 

Owing in part to these limitations in the numerics, and in
part to the fact that the initial conditions were not chosen specifically
to reproduce the structure and assembly history of the Milky Way suggested
by observational constraints, the comparison with the Milky Way
has to be done with caution. Indeed, the
simulated galaxies more closely resemble L$_*$ Sa galaxies than a Milky Way-like
Sb/Sc galaxy, as shown by their location in the Tully Fisher relation \citep{2008ASPC..396..453G}.
They have a larger bulge-to-disc ratio relative to the Milky
Way \citep{2008arXiv0801.3845M}, $B/D \sim 0.5-0.65$ (measured
both photometricaly and kinematically, see \bcite{2007MNRAS.374.1479G}; \bcite{2008ASPC..396..453G}) rather than the $0.2-0.3$ measured for the Milky Way, a more massive stellar halo
and a lower gas fraction in the disc.
As a result, while the total baryonic mass of these galaxies is comparable
to that of the Milky Way,
the stellar disc mass is a factor of 2-3 lower than that of the disc of
the Milky Way, and the gas mass in the disc is lower by an even greater
factor.
The lower disc mass combined with the effect of softening, that tends to
smear out physical density, yields a stellar volume density lower by a
factor of $\sim 3.5$ compared to the stellar density of the Milly Way at
the solar neighborhood and
a total baryonic (star+gas) density in the disc plane lower by
a factor $\sim 10$ at the same radius (\bcite{2000MNRAS.313..209H};
and see Figure \ref{fig:cosmo2}(i-l)). Due to the same structural
differences, in particular to the larger bulge-to-disc ratio, the
mean streaming velocity also falls with radius faster than in the Milky
Way (Figure \ref{fig:cosmo2}(a-d)).
This will cause disc-plane dragging -- and therefore the mass of the dark
disc -- to be underestimated\footnotemark. 

\footnotetext{There is a competing effect that has the opposite sign: the merging satellites are themselves over-concentrated due to the same numerical limitations. This causes the dark matter, through adiabatic contraction, to be over-concentrated also. This has two effects. Firstly the satellites are more resilient to tides and more able to get down into the disc. Secondly, there is more dark matter at small radii within the satellites. Both effects will act to overestimate the dark disc mass. However, even without any baryons, \MWthreedark\ has significant contributions from its accreted satellites at small radii (compare Figures \ref{fig:cosmo2}(c) and (d), dotted and dashed lines). This suggests that this effect is small.} 

Despite these numerical and observational limitations, our qualitative results are robust. This is because dark disc formation is a gravitational process that requires just three key ingredients to be correctly modelled: (i) having a stellar/gas disc in place at high redshift as observed in real galaxies; (ii) having a dark matter halo that is oblate and aligned with the stellar disc; and (iii) having a merger history given by our current $\Lambda$CDM cosmology. All three are satisfied by our simulations. 

With the above limitations in mind, we tentatively suggest that the Milky Way is intermediate between simulations \MWone\ and \MWtwo. Future simulations that better resolve the Milky Way and its satellites and that are tuned to reproduce the Milky Way's merger history will provide improved constraints.

\section{Conclusions}\label{sec:conclusions}

Predicting the local phase space density of dark matter is central to efforts to directly detect dark matter, both to motivate detector design and to interpret any future signal. Previous efforts to estimate this have used simulations that model the dark matter alone. In this paper, we used three state-of-the-art $\Lambda$CDM cosmological hydrodynamic simulations of Milky Way mass galaxies to include the stars and gas self-consistently for the first time. Our main findings are as follows: 

\begin{enumerate}

\item Once the stars and gas are included, the dark matter halos become significantly rounder, while the stellar/gas disc biases the accretion of massive satellites, dragging them towards the disc plane. Both effects combined lead to the formation of a dark matter disc with $\sim 0.25 - 1.5$ times the halo density at the solar neighbourhood, in excellent agreement with a previous estimate obtained using different methodology \citep{2008MNRAS.389.1041R}. The resulting solar neighbourhood dark matter velocity distributions are anisotropic, and better-fit in $v_\phi$ by a double Gaussian than a single Gaussian. The highly rotating second Gaussian component is the `dark disc'. We found a rotation lag compared to the Milky Way stellar disc in the range: $v_\mathrm{lag} \sim 0 - 150$\,km/s. 

\item Our three Milky Way mass galaxies were chosen to span a range of interesting merger histories, not to be precise replicas of the Milky Way. We found that more late mergers gave rise to a more significant dark disc. The Milky Way likely has a dark disc intermediate between our most quiescent galaxy, and our second most quiescent galaxy. This suggests a local dark disc density of $\rhodd/\rhoh \sim 0.23 - 1$ and rotation lag with respect to the Milky Way's stellar disc of $v_\mathrm{lag} = 0 - 150$\,km/s. Increased resolution and cosmological models that better capture the structural properties of the Milky Way will give improved constraints. For median values of $\rhodd/\rhoh \sim 0.5$ and $v_\mathrm{lag} \sim 50$\,km/s, the dark disc boosts WIMP capture in the Earth and Sun (\bcite{2008arXiv0811.4172B}; \bcite{bruchinprep}), enhances the annual modulation signal, and leads to distinct variations in the flux as a function of recoil energy that allow the WIMP mass to be determined \citep{2008arXiv0804.2896B}.

\item An accreted disc of stars forms concurrent with the dark disc and shares similar kinematics. If future surveys of our Galaxy can disentangle accreted stars in the Milky Way thick disc from those that formed in-situ, then we will be able to infer, through numerical modelling, the velocity distribution function of the dark disc from these stars. We will investigate this further in future work. 

We conclude that it is vital to model the baryons when calculating the local phase space distribution of dark matter in the Milky Way.

\end{enumerate}

\section{Acknowledgements}
J.R. would like to acknowledge support from a Forschungskredit grant from the University of Z\"urich; L.M. from SNF grant PP0022-110571; and F.G. from a Theodore
Dunham grant, HST GO-1125, NSF ITR grant PHY-0205413
(also supporting TQ), NSF grant AST-0607819 and NASA ATP
NNX08AG84G.

\bibliographystyle{mn2e}
\bibliography{/Users/justinread/Documents/LaTeX/BibTeX/refs}

\end{document}